\documentclass[12pt]{article}
\usepackage{a4wide,graphicx,amsmath,amssymb}
\usepackage{rotating}
\usepackage[numbers,sort&compress]{natbib}

\def\GeV{{\rm GeV}}

\def\msbar{\overline{MS}}
\def\lapproxeq{\lower .7ex\hbox{$\;\stackrel{\textstyle                                                    
<}{\sim}\;$}}                                                    
\def\gapproxeq{\lower .7ex\hbox{$\;\stackrel{\textstyle                                                    
>}{\sim}\;$}} 
\begin{document}
\titlepage

\begin{flushright}
LCTS/2012-25\\
\end{flushright}

\vspace*{0.2cm}

\begin{center}
{\Large \bf  The effect on PDFs and $\alpha_S(M_Z^2)$ due to changes 
in flavour scheme and higher twist contributions}

\vspace*{1cm}
\textup{R.S. Thorne \\
Department Of Physics and Astronomy, University College London\\
          Gower Place, London, WC1E 6BT, UK\\ 
        E-mail: robert.thorne@ucl.ac.uk}
\vspace*{0.2cm} 

\end{center}

\vspace*{0.2cm}

\begin{abstract}
I consider the effect on MSTW partons distribution functions (PDFs)
due to changes in the choices of theoretical procedure used in the fit. 
I first consider using the 3-flavour fixed flavour number scheme instead 
of the standard general mass variable flavour number scheme used in the 
MSTW analysis. This results in the light quarks increasing at all relatively 
small $x$ values, the gluon distribution becoming smaller at high values of 
$x$ and larger at small $x$, the preferred value of the coupling constant 
$\alpha_S(M_Z^2)$ falling, particularly at NNLO, and the fit quality 
deteriorates. I also consider lowering the kinematic
cut on $W^2$ for DIS data and simultaneously 
introducing higher twist terms which are fit
to data. This results in much smaller effects  
on both PDFs and
$\alpha_S(M_Z^2)$ than the scheme change, except for quarks at very high $x$. 
I show that the structure function one obtains from a fixed input set of PDFs 
using the fixed flavour scheme and variable flavour scheme differ
significantly for $x \sim 0.01$ at high $Q^2$, and that this is due 
to the fact that in the fixed flavour scheme there is a slow convergence 
of large logarithmic terms of the form $(\alpha_S\ln(Q^2/m_c^2))^n$ 
relevant for this 
regime. I conclude that some of the most significant differences in PDF
sets are largely due to the choice of flavour scheme used.
\end{abstract}

\vspace*{0.2cm}

\section{Introduction}

There have recently been various improvements in the PDF determinations by 
the various groups (see e.g. \cite{Martin:2012da,Ball:2012cx,Gao:2013xoa,
CooperSarkar:2011aa,Alekhin:2013nda,JimenezDelgado:2012zx}) 
generally making the predictions using different PDF sets more consistent
with each other. However, there still remain some large differences which are 
occasionally much bigger than the individual PDF uncertainties 
\cite{Watt:2011kp,Forte:2013wc,Ball:2012wy}. This is particularly the case 
for cross sections depending on the high-$x$ gluon or on higher powers of the 
strong coupling constant $\alpha_S$. In this article I investigate 
potential reasons for these differences, based on alternative theoretical 
procedures that can be chosen for a PDF fit. The two main potential sources of 
differences which may affect rather generic features such as the general
form of the gluon distribution and the preferred value of $\alpha_S(M_Z^2)$,
(rather than more detailed features such as quark flavour decomposition), are
the choice of active flavour number used and whether or not higher twist 
corrections are applied to theory calculations, and related to this whether 
low $Q^2$ and $W^2$ data are used in a PDF fit. I discover that the issue of
heavy flavours is by far the more important of these, and explain the 
reason why the differences between PDFs obtained using fixed flavour number 
scheme (FFNS) and those using a general mass variable flavour number scheme
(GM-VFNS) is so great at finite order in perturbative QCD.     
This study builds on some initial results in \cite{Thorne:2012az}
and in many senses is similar to the NNPDF study 
in \cite{Ball:2013gsa} and 
reaches broadly the same conclusions. However, there are a variety of 
differences to the NNPDF study, 
not least the investigation of the $\alpha_S$ dependence, and also 
a much more detailed discussion of the  theoretical understanding of 
the conclusions. A very brief summary of the results here have been presented in 
\cite{Thorne:2013hpa}.

\section{Flavour Number}

I first examine the number of active quark flavours used in the 
calculation of structure functions. There are essentially two 
different choices for how one deals with the charm and bottom quark
contributions, the former being of distinct phenomenological importance 
as the charm contribution to the total $F_2(x,Q^2)$ at HERA can be of 
order $30\%$. Hence, I will concentrate on the charm contribution to 
structure functions $F^c(x,Q^2)$, but all theoretical considerations are 
the same for the bottom quark contribution. 
In the $n_f=3$ Fixed Flavour Number Scheme (FFNS)
we always have 
\begin{equation}
F^c(x,Q^2)=C^{FF,c, 3}_k(Q^2/m_c^2)\otimes f^{3}_k(Q^2),
\end{equation} 
i.e. for $Q^2\sim m_c^2$ massive quarks are only  
created in the final state.
This is exact (up to nonperturbative corrections) 
but does not sum $\alpha_S^n \ln^n Q^2/m_c^2$ terms in the 
perturbative expansion. The FFNS has long been fully known at NLO 
\cite{Laenen:1992zk}, but this is not yet the case at 
NNLO (${\cal O}(\alpha_S^3)$).
Approximate results can be derived \cite{Kawamura:2012cr}, 
and are sometimes used in fits, e.g. \cite{Alekhin:2012ig}).      
However, it turns out that these NNLO corrections are not 
actually very large, except near threshold and at very low $x$, 
being generally 
of order $10\%$ or less away from these regimes. (Perhaps surprisingly, 
the approximate NNLO corrections also do not reduce the scale dependence 
by much compared to NLO, see e.g. 
Figs. 12 and 13 of \cite{Kawamura:2012cr}.) Hence, the use of 
approximate NNLO corrections to $F^c(x,Q^2)$ has not led to significant 
changes compared to NNLO PDFs which used the simpler approximation of 
only going to NLO in $F^c(x,Q^2)$, e.g \cite{Alekhin:2009ni}.  

\begin{figure}[htb!]
\centerline{\hspace{-0.7cm}
\includegraphics[width=0.45\textwidth]{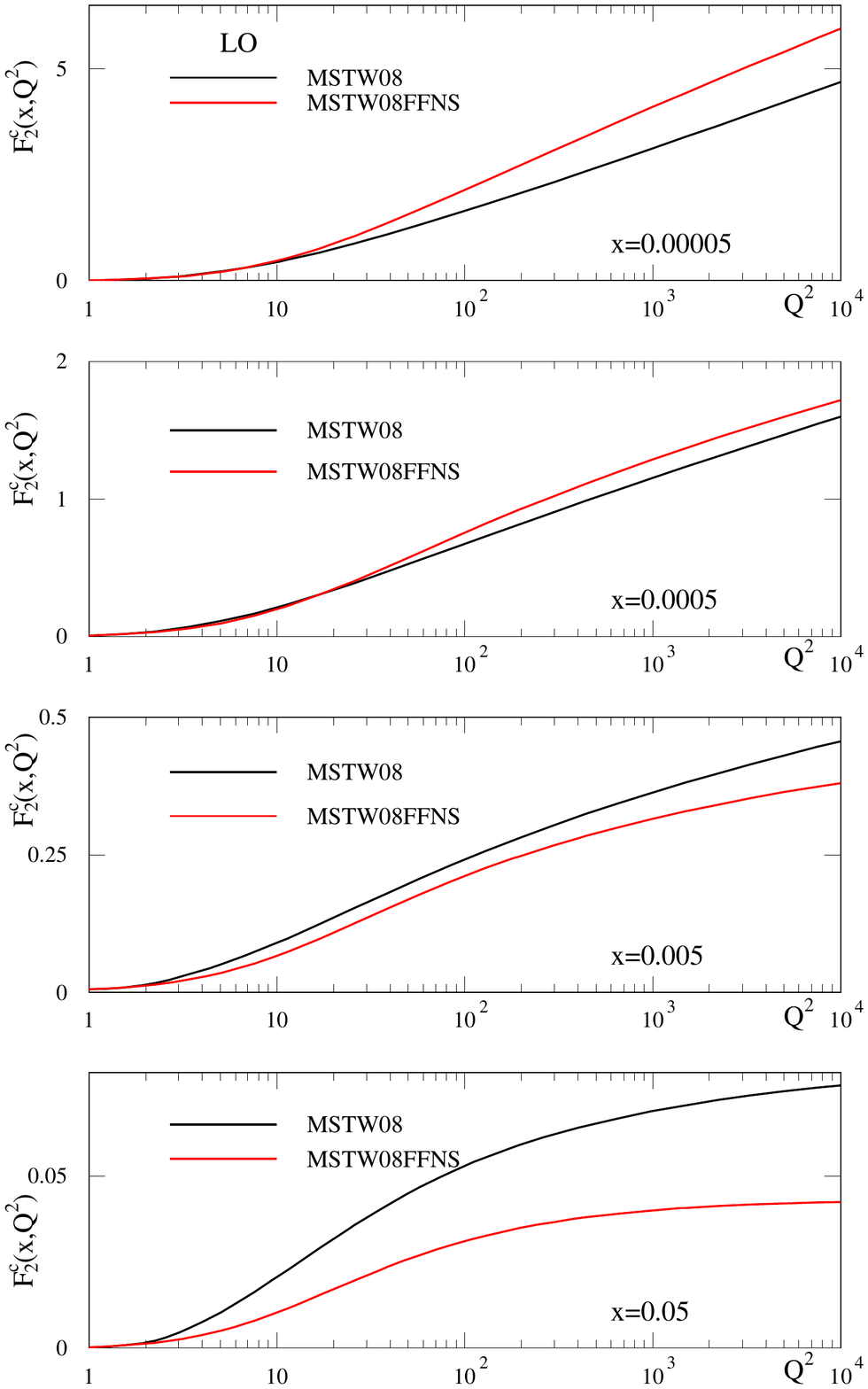}
\hspace{-1.8cm}\includegraphics[width=0.45\textwidth]{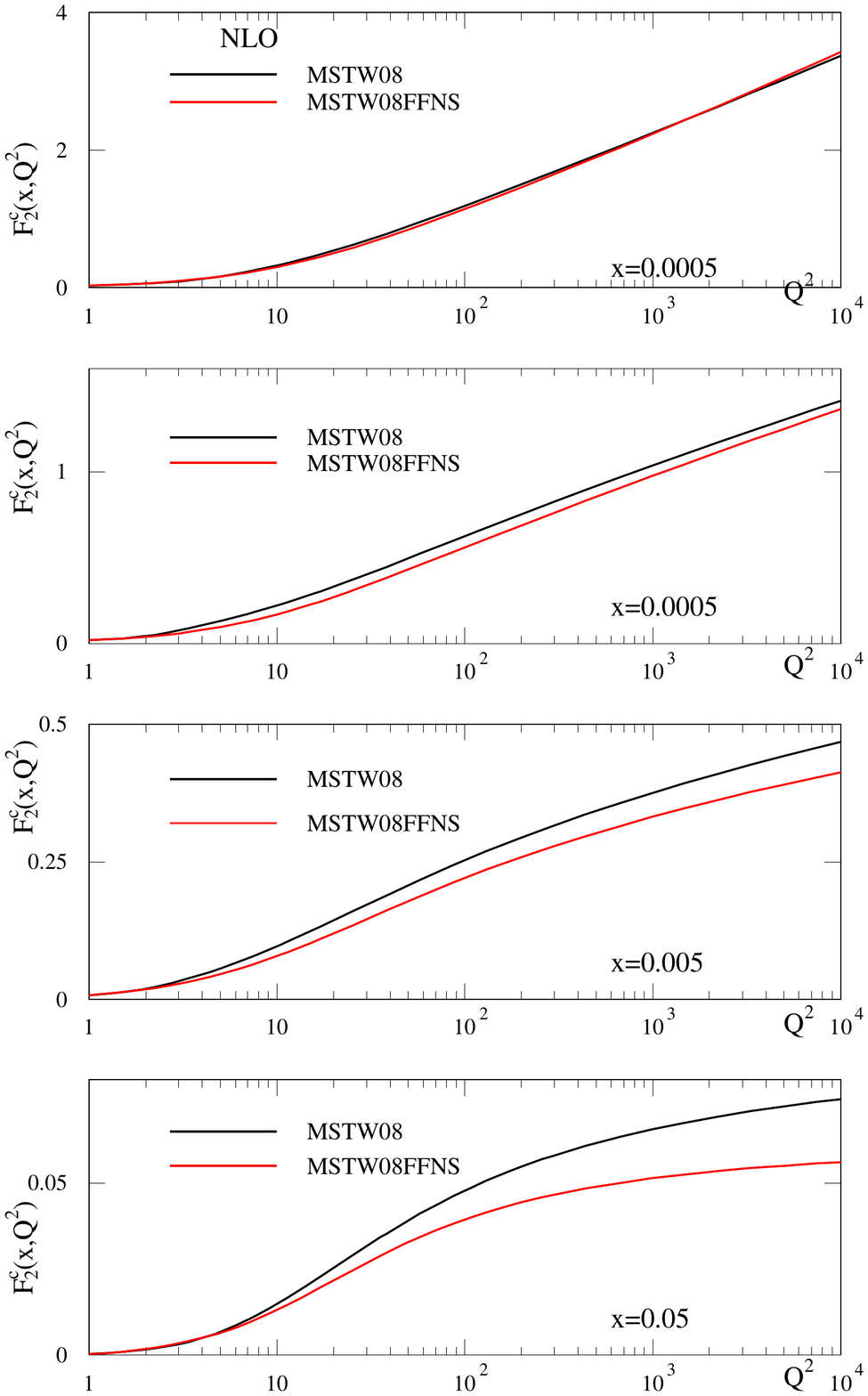}
\hspace{-1.8cm}\includegraphics[width=0.45\textwidth]{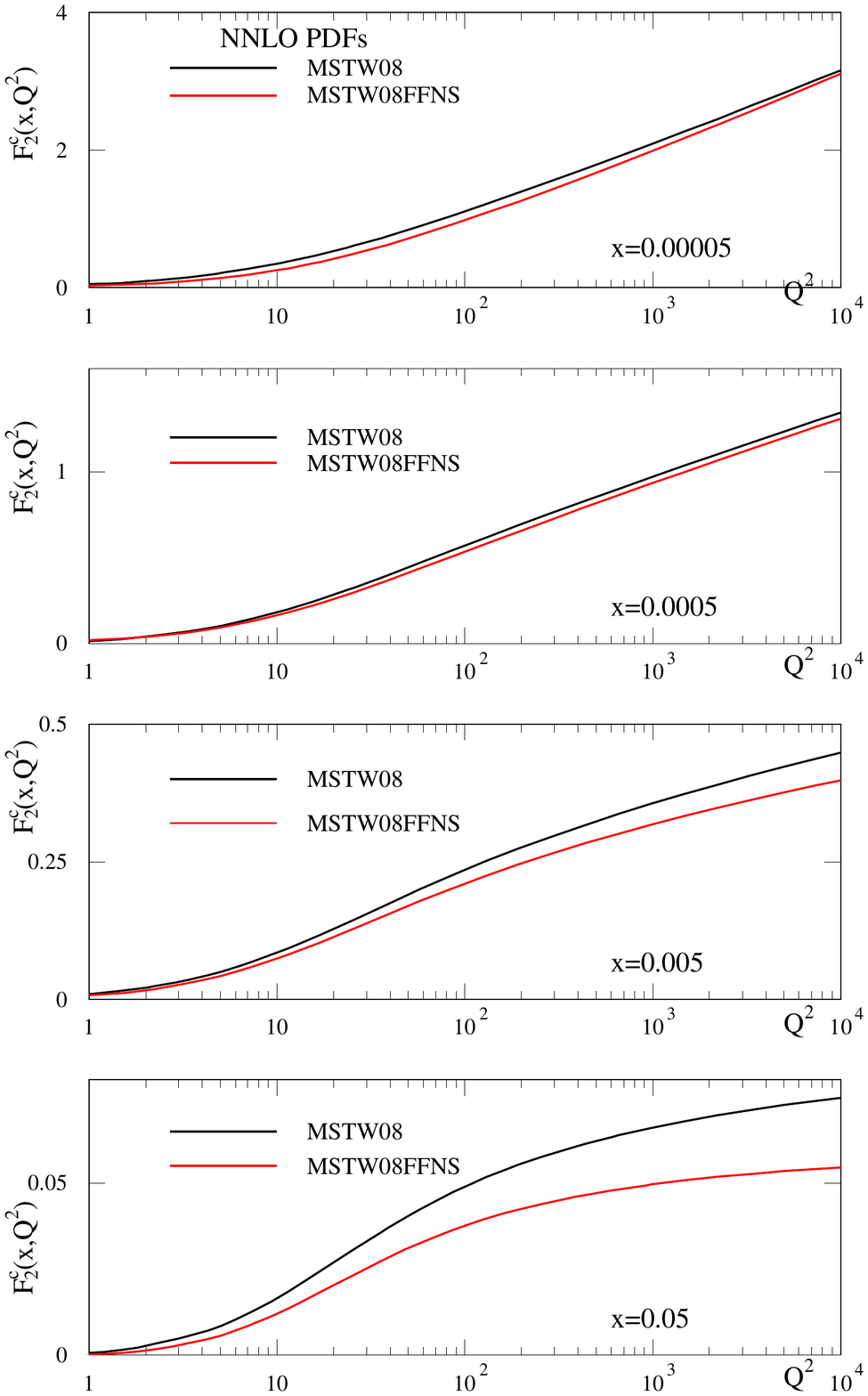}
\hspace{-1.5cm}}
\vspace{-0.7cm}
\caption{$F_2^c(x,Q^2)$ using the FFNS and GM-VFNS at LO, NLO and NNLO.
${\cal O}(\alpha_S^2)$ coefficient functions are used for FFNS at NNLO.}
\label{Fig1} 
\end{figure}

In a  variable flavour scheme  one uses the fact that 
at $Q^2 \gg m_c^2$ 
the heavy quarks behave like massless partons and the   
$\ln(Q^2/m_c^2)$ terms are automatically summed via 
evolution. PDFs in different number regions are related 
perturbatively, 
\begin{equation}
f^{4}_j(Q^2)= A_{jk}(Q^2/m_c^2)\otimes f^{3}_k(Q^2),
\end{equation} 
where the
perturbative matrix elements $A_{jk}(Q^2/m_c^2)$
are known exactly to NLO \cite{Buza:1995ie,Buza:1996wv}.\footnote{NNLO
contributions are being calculated \cite{Bierenbaum:2007qe}-\cite{Ablinger:2014lka} and are used in the approximate NNLO expressions for $F_2^c(x,Q^2)$  
in  \cite{Kawamura:2012cr}.}
The original Zero Mass Variable Flavour 
Number Scheme (ZM-VFNS)
ignores all ${\cal O}(m_c^2/Q^2)$ corrections in cross sections, i.e. 
for structure functions 
\begin{equation}
F(x,Q^2) = C^{ZM,4}_j\otimes f^{4}_j(Q^2),
\end{equation}
but this is an approximation at low $Q^2$. The majority of PDF groups use
a  General-Mass Variable Flavour Number Scheme 
(GM-VFNS). 
This is designed to take one from the well-defined limits of 
$Q^2\leq m_c^2$ where the FFNS description applies to $Q^2\gg m_c^2$ 
where the variable flavour number description is more applicable in a 
well defined theoretical manner. Some of the variants are reviewed and 
compared in \cite{Binoth:2010ra}, and for specific examples see 
e.g. \cite{Aivazis:1993pi,Thorne:1997ga,
Chuvakin:1999nx,Tung:2001mv,Thorne:2006qt,Forte:2010ta} 
There is an ambiguity in precisely how one defines 
a GM-VFNS at fixed order in perturbation theory (in the same way there is 
a renormalisation and factorisation scale uncertainty), but this is always 
formally higher order than that at which one is working.  
A study of the variation of both $F^c(x,Q^2)$ and extracted PDFs was made in 
\cite{Thorne:2012az}, and both reduced significantly at NNLO. PDFs 
and predictions for LHC cross sections could 
vary by amounts of order the experimental PDF uncertainty at NLO, i.e. $\sim 2\%$
but this reduced to generally fractions of a percent at NNLO. In both cases 
there was little variation in the preferred values of $\alpha_S(M_Z^2)$.  
Some results of variations in GM-VFNS definition can also be found in 
\cite{Gao:2013wwa}.

\begin{figure}[htb!]
\centerline{\includegraphics[width=0.6\textwidth]{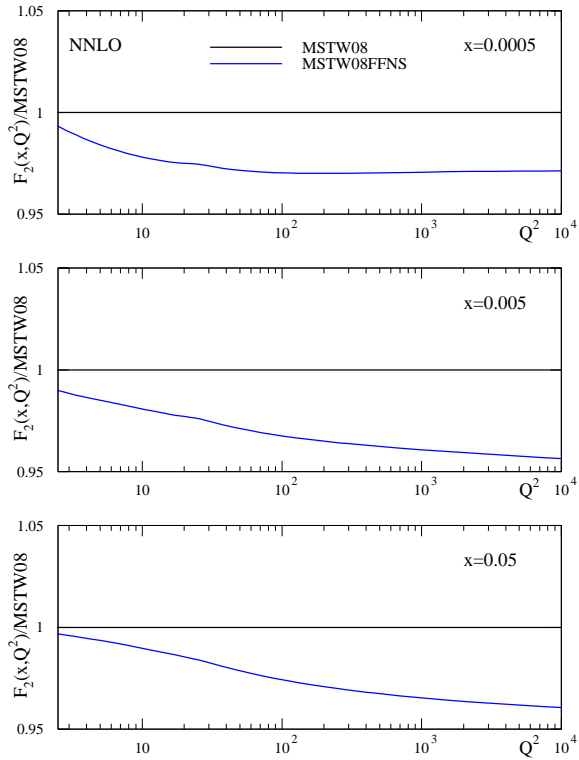}}
\vspace{-1.7cm}
\caption{The ratio of $F_2(x,Q^2)$ using the FFNS to that using the GM-VFNS.}
\label{Fig2} 
\end{figure}

The predictions for $F_2^c(x,Q^2)$ using the TR' GM-VFNS \cite{Thorne:2006qt}
and the MSTW2008 PDFs \cite{Martin:2009iq}
are compared to those using the FFNS and three-flavour PDFs generated using 
the MSTW2008 input distributions \cite{Martin:2010db}, and are shown in 
Fig.~\ref{Fig1}. At LO there is a very big difference between the two, 
particularly for $x \sim 0.05$ where the GM-VFNS result is larger than 
the FFNS result, but also at very low $x$ where the FFNS is larger. 
At NLO  $F_2^c(x,Q^2)$ at high $Q^2$ for the FFNS is nearly always lower than 
for the GM-VFNS, significantly so at higher $x\sim 0.05$. For FFNS at 
NNLO only NLO coefficient functions are used, but (various choices of)   
approximate ${\cal O}(\alpha_S^3)$ corrections
give only small increases that would not change the plots in any qualitative manner. There is no dramatic improvement in the agreement between 
FFNS and GM-VFNS at NNLO compared to NLO, contrary to what one might expect. 
This suggests that logarithmic terms beyond 
${\cal O}(\alpha_S^3 \ln^3(Q^2/m_c^2))$ are still important.  

\begin{table}[h]
\begin{center}
\begin{tabular}{|l|c|c|c|c|} \hline
NLO &  &  & &\\
\hline 
& $\chi^2$ DIS  & $\chi^2$ ftDY   & $\chi^2$ jets & $\alpha_S^{n_f=5}(M_Z^2)$\\
& 2073pts  & 199pts  & 186pts& \\
\hline
MSTW2008    & 1876 & 242 & 170 & 0.1202 \\ 
MSTW2008 (DIS only) & 1845 &  & (193) & 0.1197 \\ 
MSTW2008 (no jets) & 1875 & 241 & (181) & 0.1973\\
 \hline
MSTW$n_f=3$ (DIS only) & 1942 &  & ($>$300) & 0.1187\\ 
MSTW$n_f=3$ (DIS + ftDY) & 2000 & 261 & ($>$300)& 0.1185 \\ 
MSTW$n_f=3$ (jets) & 2010 & 269 & 177 & 0.1222\\ 
MSTW$n_f=3$ (jets+$Z$) & 2062 & 258 & 177 & 0.1225\\ 
\hline
\end{tabular}
\caption{\label{tab:t1} The $\chi^2$ values for DIS data, fixed target 
Drell Yan (ftDY) data and Tevatron jet data for various NLO fits performed using the 
GM-VFNS used in the MSTW 2008 global fit and using the $n_f=3$ FFNS 
for structure functions. The bracketed numbers denote the $\chi^2$ values for 
jet data when not included in the fit.}
\end{center}
\end{table}

\begin{table}[h]
\begin{center}
\begin{tabular}{|l|c|c|c|c|} \hline
NNLO  & &  & &\\
\hline 
& $\chi^2$ DIS  & $\chi^2$ ftDY  & $\chi^2$ jets & $\alpha_S^{n_f=5}(M_Z^2)$\\
& 2073pts & 199pts & 186pts& \\
\hline
MSTW2008    & 1864 & 251 & 177 & 0.1171 \\ 
MSTW2008 (DIS only) & 1822 &  & (292) & 0.1155 \\ 
MSTW2008 (no jets) & 1855 & 250 & (298)& 0.1160 \\
 \hline
MSTW$n_f=3$ (DIS only) & 2003 &  & ($>$300) & 0.1144\\ 
MSTW$n_f=3$ (DIS + ftDY) & 2032 & 254 & ($>$300)& 0.1152 \\ 
MSTW$n_f=3$ (jets) & 2094 & 270 & 179 & 0.1181\\ 
MSTW$n_f=3$ (jets+$Z$) & 2172 & 258 & 179 & 0.1184\\ 
\hline
\end{tabular}
\caption{\label{tab:t2} The $\chi^2$ values for DIS data, fixed target 
Drell Yan (ftDY) data and Tevatron jet data for various NNLO fits performed using the 
GM-VFNS used in the MSTW 2008 global fit and using the $n_f=3$ FFNS 
for structure functions. The bracketed numbers denote the $\chi^2$ values for 
jet data when not included in the fit.}
\end{center}
\end{table}

This $20$-$40\%$ difference between FFNS and GM-VFNS in $F_2^c(x,Q^2)$
can lead to over $4\%$ changes in the total inclusive structure function 
$F_2(x,Q^2)$, see Fig.~\ref{Fig2} for an illustration at NNLO, 
with the GM-VFNS result usually being above the FFNS result. 
At $x\sim 0.01$ this is mainly due to the difference in $F_2^{c}(x,Q^2)$
itself. However, at lower $x$ there is a contribution to the difference from 
the light quarks evolving slightly more slowly in the FFNS, mainly due to 
the strong coupling in the FFNS falling below that in the GM-VFNS as 
$Q^2$ increases above $m_c^2$.   
For $x>0.1$ the FFNS and GM-VFNS are very similar largely because the 
charm contribution is becoming very small, and the valence quark 
contribution dominates. In order to test the importance of 
this difference between FFNS and GM-VFNS 
in inclusive $F_2(x,Q^2)$ I have extended an investigation 
begun in \cite{Thorne:2012az} and performed fits using
the FFNS scheme in order to compare the fit quality and resulting PDFs
and $\alpha_S(M_Z^2)$ to those obtained from fits using the GM-VFNS. 
At NNLO ${\cal O}(\alpha_S^2)$ heavy flavour coefficient functions are 
used as default (which has been done until quite recently in other 
FFNS fits, e.g. \cite{Alekhin:2009ni}). It has been checked, however, that 
approximate ${\cal O}(\alpha_S^3)$ expressions change the 
results very little.

In order to make comparison to the existing MSTW2008 PDFs, 
which have been very extensively used in LHC studies,  
I perform the fits within the framework of the MSTW2008 PDFs
\cite{Martin:2009iq}, i.e. data sets and treatment are the same, as is the 
definition of the GM-VFNS, quark masses, {\it etc.}. 
(The effect on the MSTW2008 PDFs due to numerous improvements in both 
theory and inclusion of new data sets 
(see \cite{Thorne:2010kj,Martin:2012da,Watt:2013oha}) has been studied 
and so far only received corrections of any real significance in the small-$x$ 
valence quarks from the improved parameterisation and deuteron corrections
in \cite{Martin:2012da}.)
For the fixed target Drell-Yan data the contribution of heavy flavour is 
negligible, and has been omitted in the FFNS fits. 
This study also maintains continuity with the previous results in 
\cite{Thorne:2012az}. I first perform fits to only DIS and fixed target 
Drell-Yan data (charged current HERA DIS data is omitted due to 
the absence of full ${\cal O}(\alpha_S^2)$ calculations for these
\footnote{There is a very recent calculation of the ${\cal O}(\alpha_S^2)$ 
results for charm production in the large $Q^2$ limit \cite{Blumlein:2014fqa}.}, 
though these run I data carry very little weight in the fit), 
but this is also extended to the additional inclusion of 
Tevatron jet and $Z$ boson production data, where the 5-flavour 
calculation scheme is used in these cases, with the PDFs 
being converted appropriately for combination with these hard cross sections.
At NNLO the fit to Tevatron jet data uses the NNLO threshold corrections
that are available \cite{Kidonakis:2000gi} (though more complete calculations 
which take into account the dependence on the jet radius $R$ 
have just appeared in \cite{deFlorian:2013qia} these are not available for 
use yet). As argued in \cite{Watt:2013oha} the precise form of these is not 
very important to the results.

The results of the fit quality for various different fits are shown in Table
\ref{tab:t1} for NLO and Table \ref{tab:t2} for NNLO, along with the value of 
$\alpha_S(M_Z^2)$, evaluated for 5 quark flavours. 
The fit quality for DIS and Drell-Yan 
data are at least a few tens of units higher in $\chi^2$ in the FFNS fit 
than in the MSTW2008 fit, with the difference being greater at NNLO 
than at NLO. The results appear similar to those in Table 1 of 
\cite{Ball:2013gsa}, though there $\alpha_S(M_Z^2)$ was kept fixed. 
The FFNS fit is often slightly better for the 
$F_2^c(x,Q^2)$ itself, but the total $F_2(x,Q^2)$ is
flatter in $Q^2$ for $x \sim 0.01$, and this worsens the fit to HERA 
inclusive structure function data. For both GM-VFNS and FFNS, and at both 
NLO and NNLO, the fit quality to DIS data deteriorates by about 30 units
when the fixed target Drell Yan data is added, showing that there is some 
tension in quark-antiquark decomposition between DIS and fixed-target Drell 
Yan data. Although there is no difficulty in obtaining a good fit to Tevatron 
jet data when using the the FFNS for structure functions 
the fit quality for DIS and Drell Yan deteriorates by
$\sim 50$ units when both Tevatron jet and $Z$ 
data are included, as opposed to $10$ units or less when using a GM-VFNS.
It is important to add the Tevatron $Z$ rapidity data as well 
as the jet data since the former fixes the luminosity at the Tevatron quite 
precisely, and makes the jet data more difficult to fit than when the 
luminosity is left free \cite{Thorne:2011kq} and vector boson production
ignored. The preferred 
$\alpha_S(M_Z^2)$ values in each fit are also shown. These do not vary much 
for the GM-VFNS fits, though for DIS only fits there is in fact very little 
variation in fit quality with a wide range of $\alpha_S(M_Z^2)$ and it is 
quite difficult to obtain a definite best fit. For the FFNS fits there is a 
very distinct increase when Tevatron jet data is added.  
The values of $\alpha_S(M_Z^2)$ are lower than for the GM-VFNS fits 
for the DIS and DIS plus Drell Yan fits, but higher when the jet data is 
added, though the NNLO FFNS values are relatively slightly lower compared
to GM-VFNS than the NLO values.

\begin{figure}[htb!]
\vspace{-0.8cm}
\centerline{\includegraphics[width=0.49\textwidth]{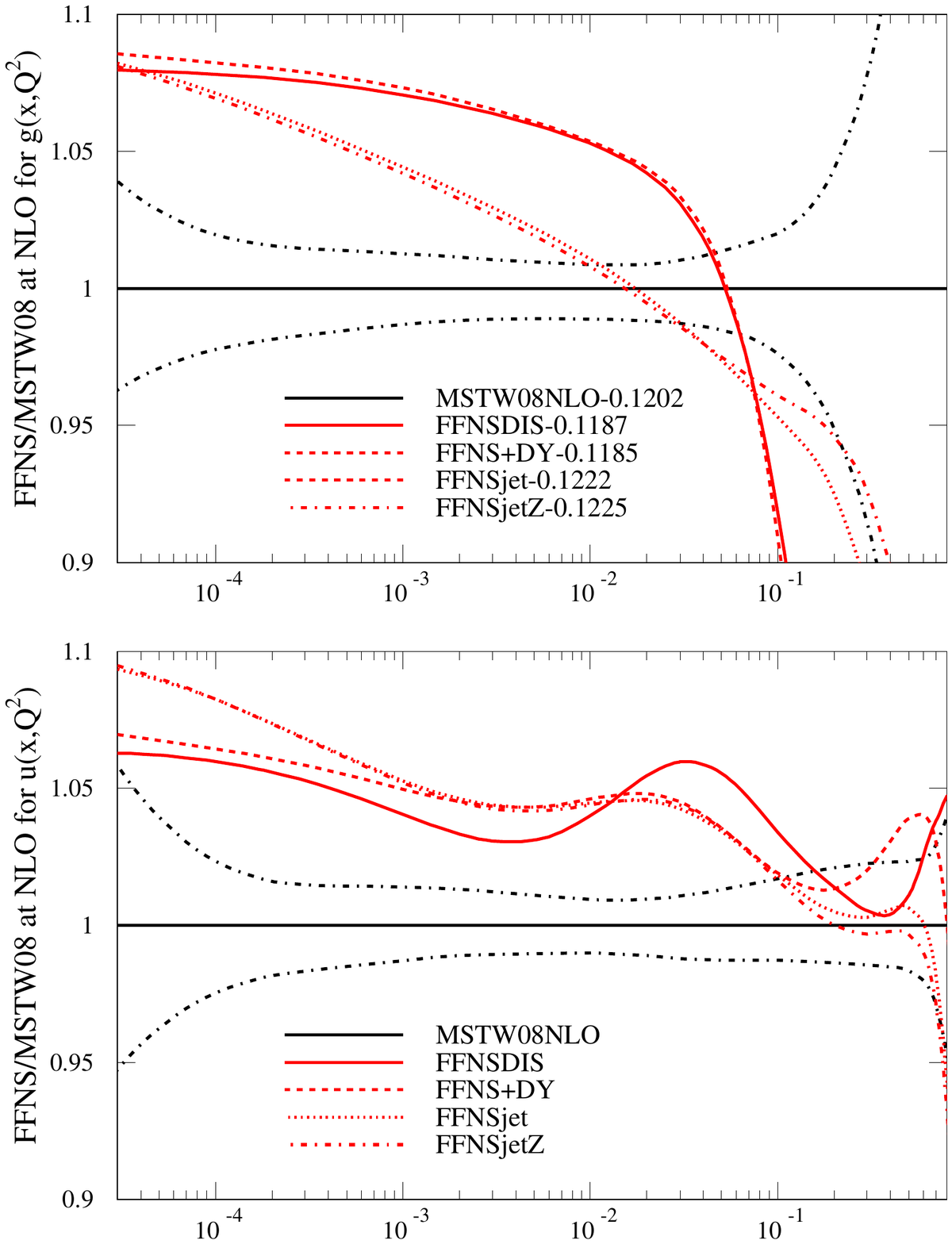}
\hspace{-1.5cm}\includegraphics[width=0.49\textwidth]{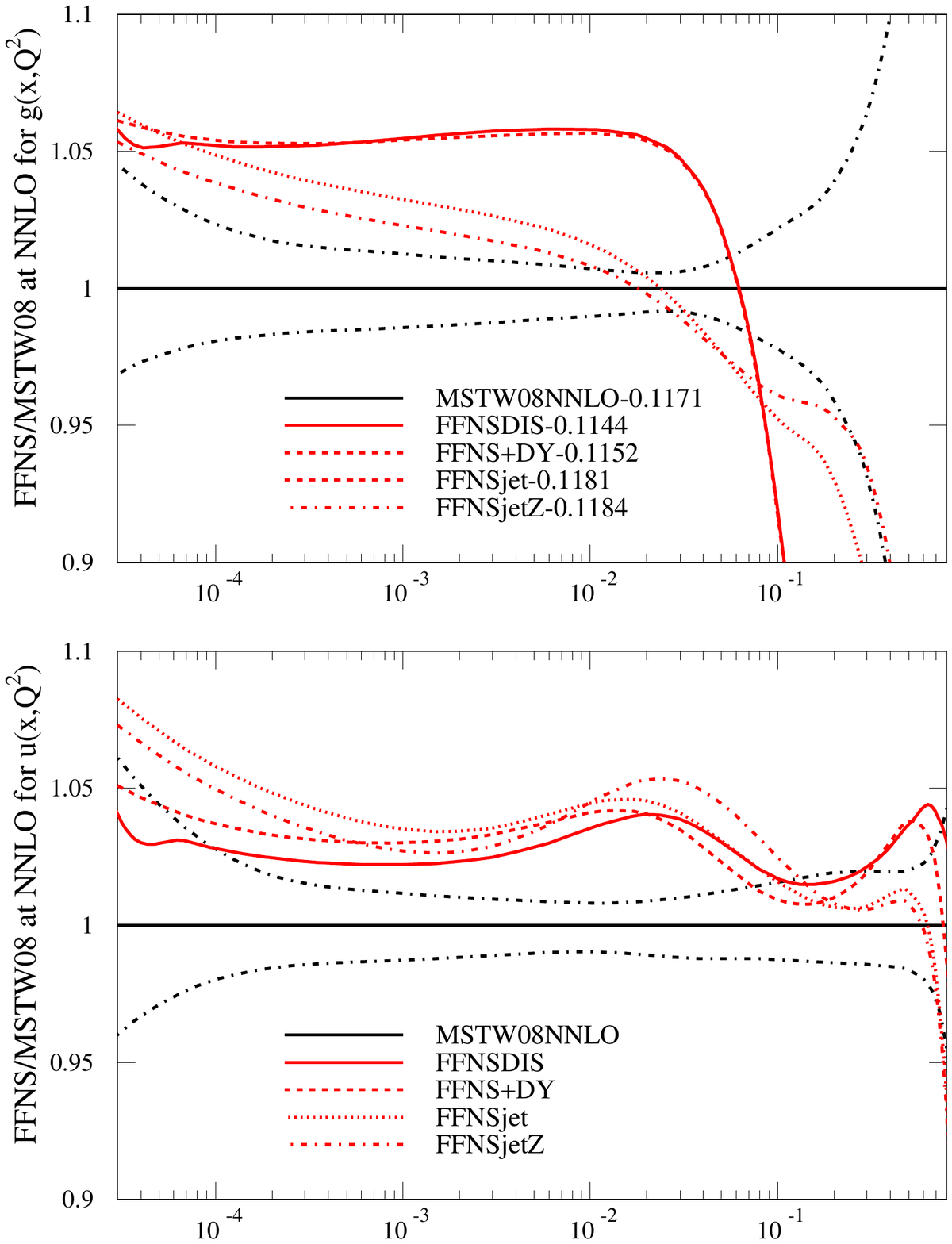}}
\vspace{-1.2cm}
\caption{Ratios of PDFs in various FFNS fits to the MSTW2008 PDFs
at $Q^2=10,000\GeV^2$.}
\label{Fig3} 
\end{figure}

\begin{figure}[htb!]
\vspace{-0.8cm}
\centerline{\includegraphics[width=0.49\textwidth]{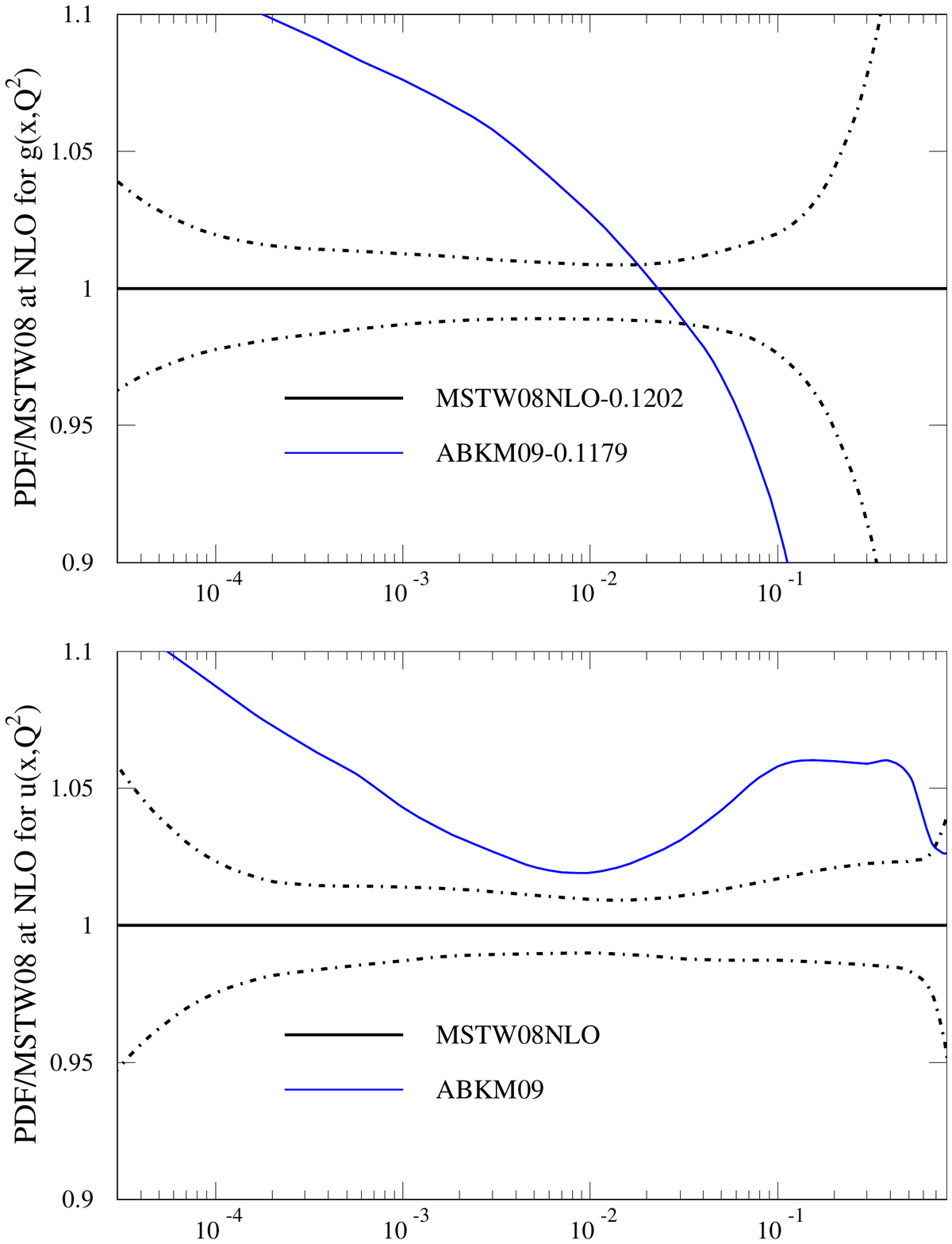}
\hspace{-1.5cm}\includegraphics[width=0.49\textwidth]{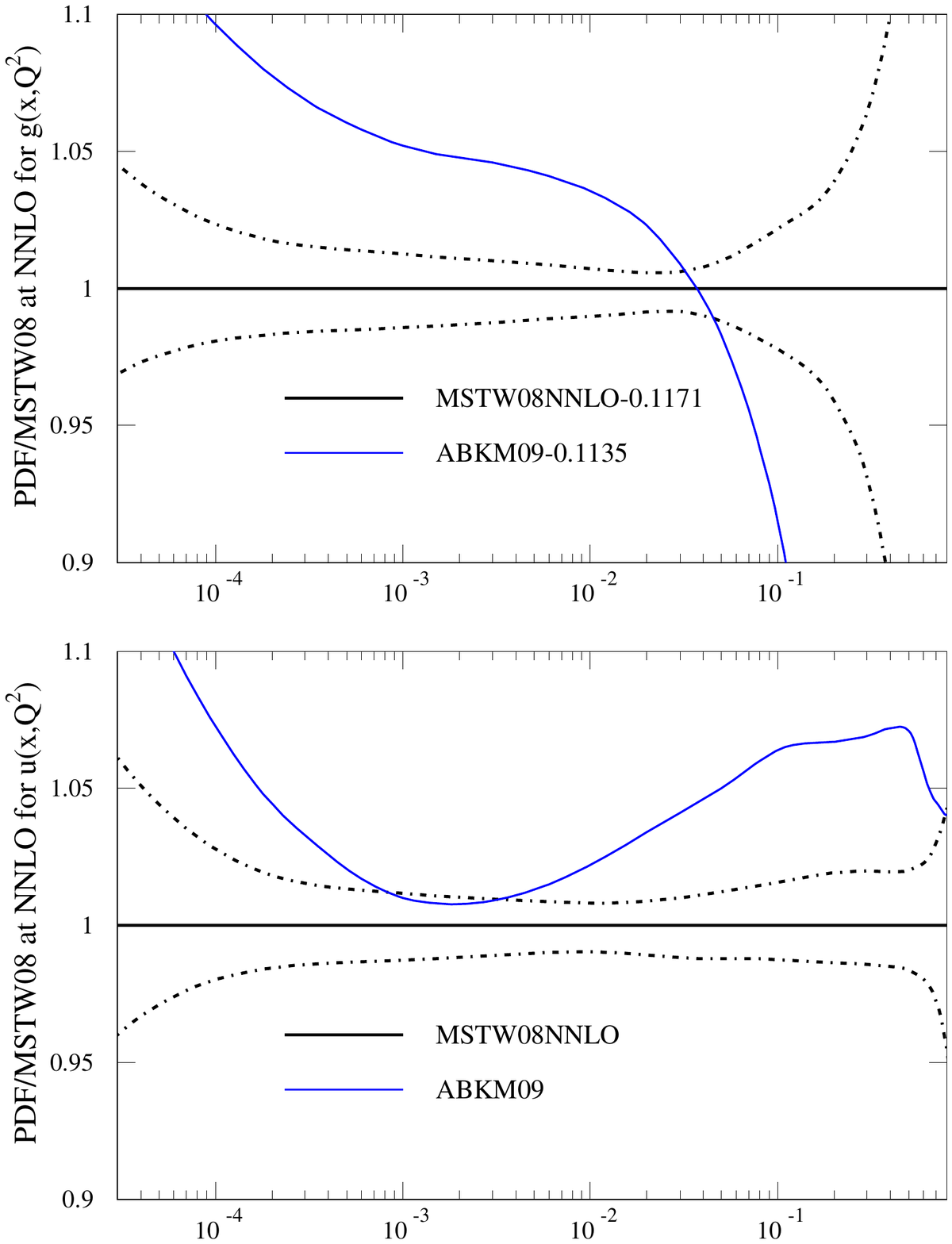}}
\vspace{-1.2cm}
\caption{Ratios of the ABKM09 PDFs to the MSTW2008 PDFs
at $Q^2=10,000\GeV^2$.
Data taken from \cite{PDFplotter}.}
\label{Fig4} 
\end{figure}

The PDFs resulting from the fits, evolved up to $Q^2=10,000\GeV^2$ (using 
variable flavour evolution for consistent comparison) are 
shown in Fig.~\ref{Fig3}. The PDFs are consistently different in form to 
the MSTW2008 PDFs. There are larger light quarks for all the FFNS fit 
variants, due to the need to make up for the smaller values of 
$F_2^c(x,Q^2)$ at high $Q^2$. The effect is very slightly reduced at NNLO 
compared to NLO. The FFNS  fits produce a gluon which is bigger at 
low $x$ that when using the GM-VFNS, and much smaller at high $x$. The 
effect is somewhat reduced when the Tevatron jet data is included in the 
fit, but not removed. Some similar differences have 
been noted in \cite{Ball:2013gsa}, though $\alpha_S(M_Z^2)$ 
was not left free, and also earlier in \cite{CooperSarkar:2007ny}.
Hence it is clear that using FFNS rather than GM-VFNS leads to significant 
changes in PDFs, and much larger changes than any
variation in choice of GM-VFNS \cite{Thorne:2012az}, particularly at NNLO.
In Fig.~\ref{Fig4} I show the same type of plot for a different PDF set 
obtained using FFNS for the structure function calculations, i.e. the ABKM 
set from \cite{Alekhin:2009ni}, which was obtained fitting to DIS and fixed 
target Drell-Yan data, and which obtained values of $\alpha_S(M_Z^2)$ of 
0.1179 and 0.1135 at NLO and NNLO 
respectively. I compare to this set, despite the fact that 
there have been more recent updates, since the data fit and the FFNS 
definition used at 
NNLO are most similar to the data used in the MSTW2008 fit and to the 
heavy flavour calculations used in this article. (More recent updates 
of the ABM fits have not led to very significant changes in the most striking
features of the comparison of FFNS to GM-VFNS PDFs, i.e. FFNS has larger light quarks, a
different shape gluon and lower $\alpha_S(M_Z^2)$.)  There are considerable 
additional differences between the fits of the two groups though, 
for instance the issue of higher twist, which is a topic to be discussed 
later. However, first I will explore the origin of the differences between 
the FFNS and GM-VFNS results.

\section{Perturbative Convergence of Heavy Flavour Evolution}

The fact that there is a considerable difference between the FFNS and GM-VFNS 
results for $F^c(x,Q^2)$ for some values of $x$, mainly $x \sim 0.05$ at NLO,
with little apparent improvement at NNLO, might seem surprising. It has 
generally been assumed that differences between the two flavour schemes would 
diminish quickly at higher orders, and hence thought unlikely that it could 
be a major source of difference between PDF sets. However, the results of 
the previous section, plus those in 
\cite{Thorne:2012az,Ball:2013gsa,CooperSarkar:2007ny}
demonstrate that differences are indeed significant, and the 
origin of this needs to be understood.    

\begin{figure}[htb!]
\vspace{-1cm}
\centerline{\includegraphics[width=0.6\textwidth]{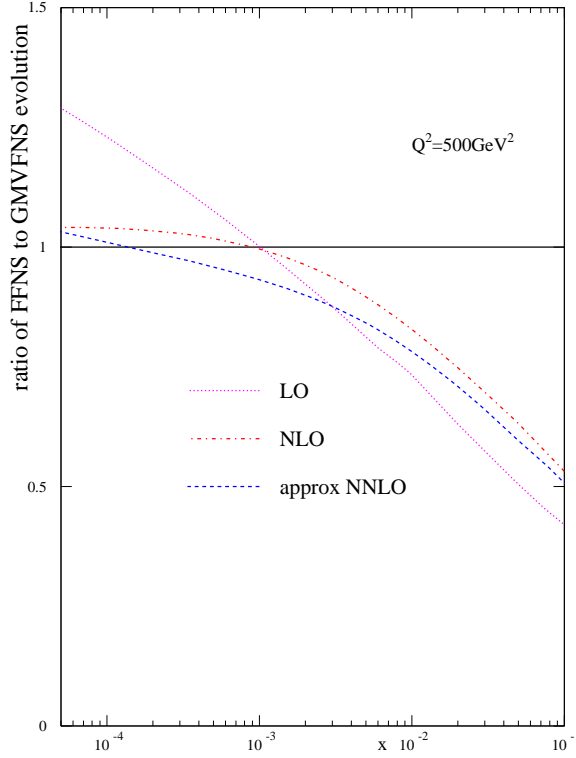}}
\vspace{-1.2cm}
\caption{The ratio of $dF_2^c/d\ln Q^2$ using the FFNS to that using the GM-VFNS at LO, NLO and NNLO.}
\label{Fig5} 
\end{figure}

In order to explain the differences between the results of 
FFNS and GM-VFNS evolution it is useful to concentrate on the relative 
size of $(dF^c_2(x,Q^2)/d \ln Q^2)$ rather than on the absolute value 
of $F^c_2(x,Q^2)$, though differences in the former clearly lead to 
differences in the latter as at very low $Q^2$ the inputs are the 
same in the two schemes. I show the ratio of $(dF^c_2(x,Q^2)/d \ln Q^2)$
in FFNS to that in GM-VFNS at LO, NLO and NNLO, using MSTW2008 PDFs, 
for $Q^2=500~\GeV^2$ in 
Fig.~\ref{Fig5}. As one can see the results mirror those for the values 
of $F_2^c(x,Q^2)$ in Fig.~\ref{Fig1} with all orders lower 
using FFNS for $x > 0.001$, but FFNS and GM-VFNS being similar at NLO and 
NNLO for very small $x$, and the LO FFNS being greater in this 
regime.\footnote{Note that these results are consistent with those in 
Fig.~5 of \cite{Alekhin:2013qqa}, which shows the difference between 
the heavy quark evolution calculated at finite order via the matrix elements 
and from full evolution. For example, at $x=0.02$ this difference is 
negative and hardly diminished at all at approximate NNLO compared to NLO. 
At lower values of $x$ the difference changes sign, but may be seen to be 
a smaller fraction of the total evolution. Exact details depend on the 
PDFs and $\alpha_S$ values used.}     
These results in the relative speed of evolution can be understood 
analytically.   

Let us begin at leading order. 
At LO in the FFNS (setting all scales to be $Q^2$, 
which is appropriate at $Q^2 \gg m_c^2$)
\begin{equation}
 F_2^{c,1,FF}  =  
\alpha_S\ln(Q^2/m_c^2) p^0_{qg}\otimes g 
+ {\cal O}(\alpha_S \cdot g) \equiv \alpha_S A_{Hg}^{1,1}
\otimes g + {\cal O}(\alpha_S \cdot g) ,
\end{equation}
where the term not involving the logarithm $\ln(Q^2/m_c^2)$ can easily be seen 
to be very sub-dominant at high $Q^2$. 
Calculating the rate of change of evolution
\begin{eqnarray}
\frac{d\,F_2^{c,1,FF}}{d \ln Q^2}  &=&  \alpha_S p^0_{qg}\otimes g
 + \ln(Q^2/m_c^2)\frac{d \, (\alpha_S p^0_{qg}\otimes g)}{d\ln Q^2} 
+ \cdots \cr
&=&  \alpha_S p^0_{qg}\otimes g
 + \ln(Q^2/m_c^2)\alpha^2_S (p^0_{qg}\otimes p^0_{gg} \otimes g
-\beta_0 p^0_{qg} \otimes g) + \cdots,
\label{FFNSloevolution}
\end{eqnarray}
where $\beta_0 =9/(4\pi) = 0.716$. A quark dependent term of 
${\cal O}(\alpha_S^2)$ (i.e. $\ln(Q^2/m_c^2)\alpha^2_S (p^0_{qg}\otimes 
p^0_{gq} \otimes \Sigma $) is deemed to be subleading. At small $x$ this is
an excellent approximation due to the smallness
of the quark distribution compared to the gluon and the fact that 
in this limit $p^0_{gq} =4/9 p^0_{gg}$. At high $x$ the quark 
distributions begin to dominate and the approximation is not as good. However,
even this is not a major issue until very high $x$, where valence quarks are 
completely dominant, since the effect of $p^0_{gq}$ is small 
compared to that of $p^0_{gg}$, e.g. the fifth moment of $p^0_{gq}$ is only 
about $-0.03$ that of $p^0_{gg}$.
    
At LO in the GM-VFNS, where $F_2^{c,1,VF}=(c + \bar c) = c^+$, to a very 
good approximation at high $Q^2$ we have
\begin{equation}
\frac{d\,F^{c,1,VF}}{d \ln Q^2} = \frac{d\,c^+}{d \ln Q^2} = \alpha_S \,p^0_{qg}\otimes g  +
\alpha_S\, p^0_{qq} \otimes c^+,
\label{cevolution}
\end{equation}
where 
\begin{equation}
c^+ \equiv \alpha_S\ln(Q^2/m_c^2) p^0_{qg}\otimes g + \cdots \equiv \alpha_S A_{Hg}^{1,1}\otimes g + \cdots 
\end{equation} 
so the second term in (\ref{cevolution}) 
is formally ${\cal O}(\alpha_S^2 \ln(Q^2/m_c^2))$.  
The first terms in Eqs. (\ref{FFNSloevolution}) and (\ref{cevolution})  
are of order $\alpha_S$ and 
they are equivalent, as they must be. 
The difference between the two LO expressions is ${\cal O}(\alpha_S^2\ln(Q^2/m_c^2))$ and is 
\begin{eqnarray}
\frac{d(F_2^{c,1,VF}\!-\!F_2^{c,1,FF})}{d \ln Q^2} &=& \alpha_S^2 \ln(Q^2/m_c^2)p^0_{qg}
\otimes (p^0_{qq} +\beta_0 -p^0_{gg}) \otimes g + \!\cdots\cr
&\equiv& P^{\rm LO}_{VF-FF} \otimes g + \!\cdots .
\end{eqnarray}  
The effect of $-p^0_{gg}$
is positive at high $x$ and negative at small $x$. That of 
$p^0_{qq}$ is negative at high $x$, but smaller than 
$p^0_{gg}$, and that of $\beta_0$ is always positive. 
Hence, the difference is large and positive at high $x$ and 
becomes large and negative at small $x$. This explains the features observed 
in Fig~\ref{Fig5}, which plots the ratio of the evolution using the FFNS 
to that using the GM-VFNS.
Hence, the difference between FFNS
and GM-VFNS evolution is fully explained.
 
The subleading terms providing the difference between FFNS and GM-VFNS
evolution at LO then provide important information about the NLO
FFNS expressions. This formally NLO difference between the two forms of 
evolution must be eliminated in the full NLO expressions by 
defining the leading-log term in the FFNS expression
to provide cancellation, i.e. it requires that  
\begin{equation}
F_2^{c,2,FF}\!\!= \alpha^2_S A_{Hg}^{2,2} \otimes g + \cdots = \frac{1}{2}
\alpha^2_S \ln^2(Q^2/m_c^2) p^0_{qg}\otimes (p^0_{qq} +\beta_0 -p^0_{gg}) \otimes g   + {\cal O}(\alpha^2_S \ln(Q^2/m_c^2)).
\end{equation}
up to quark mixing corrections and sub-dominant terms. 
With this definition all previous 
${\cal O}(\alpha_S^2\ln(Q^2/m_c^2))$ terms in the NLO evolution cancel 
between the GM-VFNS and FFNS expressions. 
However, the derivative of $F_2^{c,2,FF}$ contains 
\begin{equation}
\frac{1}{2}\ln^2(Q^2/m_c^2) \frac{d\, \bigl(\alpha^2_S p^0_{qg}\otimes (p^0_{qq} +\beta_0 -p^0_{gg}) \otimes g\bigr)}{d\, \ln Q^2}
\end{equation}
which does not cancel with anything in the NLO GM-VFNS expression. 
This leads to
\begin{equation}
P^{\rm NLO}_{VF-FF} = \frac{1}{2} \alpha_S \ln(Q^2/m_c^2)(p^0_{qq} 
+2\beta_0 -p^0_{gg}) \otimes P^{\rm LO}_{VF-FF},
\end{equation}  
where again the $p^0_{qq}$ comes form the contribution in 
Eq.~(\ref{cevolution}) but using the ${\cal O}(\alpha_S^2\ln^2(Q^2/m_c^2))$ 
contribution to $c^+$ in $\alpha^2_S A_{Hg}^{2,2} \otimes g$.  
The additional factor of $(p^0_{qq} +2\beta_0 -p^0_{gg})$ is
large and positive at high $x$ and negative at small $x$, but 
not until smaller $x$ than at LO. Therefore, $P^{\rm NLO}_{VF-FF}$
is large and positive at high $x$, negative for 
smaller $x$ and positive for extremely small $x$. This explains 
the difference in the evolution between GM-VFNS and FFNS at NLO correctly.

\begin{figure}[htb!]
\vspace{-1cm}
\centerline{\includegraphics[width=0.6\textwidth]{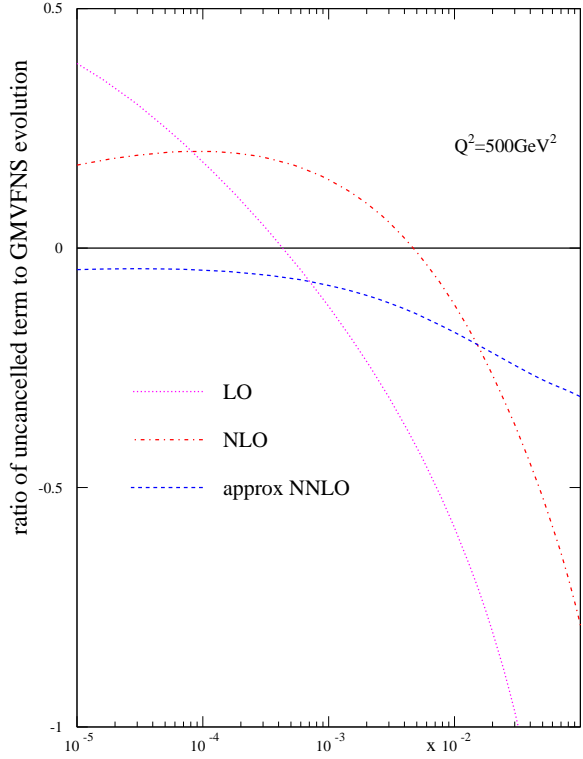}}
\vspace{-1.2cm}
\caption{The ratio of the analytic leading-log approximation to 
the evolution difference between FFNS and the full GM-VFNS evolution at 
LO, NLO and NNLO, i.e. $P_{FF-VF}\otimes g$ at each order.}
\label{Fig6} 
\end{figure}

The pattern is now established. In order to cancel this difference between 
the evolutions at NLO then at NNLO the dominant part of
$F_2^{c,2,FF}$ at leading-log is (up to quark-mixing and scheme-dependent 
terms)
\begin{equation}
\alpha^3_S A_{Hg}^{3,3}
\otimes g = \frac{1}{6}
\alpha^3_S \ln^3(\frac{Q^2}{m_c^2}) p^0_{qg}\otimes (p^0_{qq} +\beta_0 -p^0_{gg\
}) \otimes (p^0_{qq} +2\beta_0 -p^0_{gg}) \otimes g.
\end{equation}
Repeating the previous arguments, at NNLO the dominant
high-$Q^2$ uncancelled term between GM-VFNS and FFNS evolution is
\begin{equation}  
P^{\rm NNLO}_{VF-FF}=\frac{1}{3} \alpha_S \ln(Q^2/m_c^2)(p^0_{qq} +3\beta_0 -p^0_{gg})\otimes
P^{\rm NLO}_{VF-FF}.
\end{equation} 
This remains large and positive at high $x$, then changes sign 
twice but stays small until becoming negative at tiny
$x$. Again this explains the behaviour at NNLO correctly.
The expression can be straightforwardly generalised to higher orders. 
It is similar in some sense
to the results for the bottom quark of Eq.~(3.5) 
in \cite{Maltoni:2012pa}, but this neglected the evolution of the gluon and
hence the $p^0_{gg}$ terms, 
which as shown here are actually the dominant effect at lowish orders.

The extent to which these relatively simple analytic results, true at 
leading log and ignoring quark mixing, describe the 
true detailed difference between the GM-VFNS and FFNS evolution can be tested
by calculating the ratio
\begin{equation}
\frac{P^{xxLO}_{FF-VF} \otimes g}{(d\,F^{c,xxLO,VF}(x,Q^2)/d\,\ln Q^2)}
\approx
\frac{(d\,F^{c,xxLO,FF}(x,Q^2)/d\,\ln Q^2)}{(d\,F^{c,xxLO,VF}(x,Q^2)/d\,\ln Q^2)} -1 ,
\end{equation}
at LO, NLO and NNLO. With the addition of unity 
this should be the same as  the result of FFNS to GM-VFNS evolution shown in 
Fig.~\ref{Fig5}. The ratio is shown in Fig.~\ref{Fig6}. 
Indeed the comparison to Fig.~\ref{Fig5}, though not exact is 
generally very good, with the most important feature of a suppression of FFNS
evolution compared to GM-VFNS of at least $20\%$ for $x\sim 0.01$, with slow 
convergence at higher orders, explained well by the simple expression.

\begin{figure}[htb!]
\vspace{0.5cm}
\centerline{\includegraphics[width=0.7\textwidth]{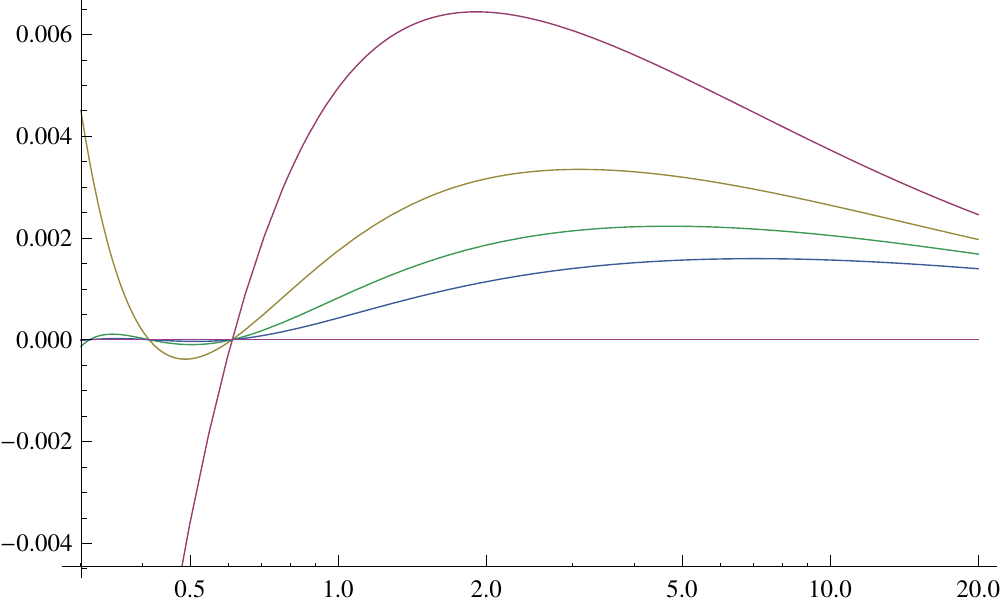}}
\caption{The effective anomalous dimension $\gamma_{VF-FF}(N)$ 
for $Q^2=500~\GeV^2$ at LO (purple), 
NLO (brown) and NNLO (green). Also shown (blue) is the NNNLO expression.}
\label{Fig7} 
\end{figure}

In order to look at the effect of this dominant high-$Q^2$ difference 
between GM-VFNS and FFNS evolution, and in particular to understand 
the rate of convergence between the two, it is 
useful to define the moment space effective anomalous dimension 
$\gamma_{VF-FF}$ obtained from from the effective splitting function 
$P_{VF-FF}$ by
\begin{equation}
\gamma_{VF-FF}(N,Q^2) = \int_0^1 x^{N} P_{VF-FF}(x,Q^2).
\end{equation}
This is shown at LO, NLO 
and NNLO for $Q^2=500\GeV^2$ in Fig.~\ref{Fig7}. Since the expression depends
only on leading logs it can actually be expressed at any order, so NNNLO is 
also shown. At high $Q^2$, values of 
$x\sim 0.05$ correspond to $N\sim 2$, where  $\gamma_{VF-FF}$ only tends to 
zero slowly as the perturbative order increases. This explains why FFNS 
evolution for $x \sim 0.05$ only slowly converges to the GM-VFNS result
with increasing order, very roughly like $1/n$ where $n$ is the power of 
$\alpha_S(Q^2)\ln(Q^2/m_c^2)$. For $N\approx 0.5$ which is applicable 
to $x\sim 0.0001$ there is good convergence, and in fact very little 
difference between FFNS and GM-VFNS evolution. For $N \to 0$, there is poor 
convergence, but this only affects extremely low values of $x$ indeed. 
It is the slow convergence relevant for $x \sim 0.05$ that is of 
phenomenological importance, as there is a great deal of very precise HERA 
inclusive structure function data that is sensitive to this.

\section{Higher Twist}

\begin{table}[h]
\begin{center}
\begin{tabular}{|c|r@{.}l|r@{.}l||r@{.}l|r@{.}l|} \hline
& \multicolumn{2}{c|}{} & \multicolumn{2}{c|}{}
& \multicolumn{2}{c|}{} & \multicolumn{2}{c|}{} \\
$x$ & \multicolumn{2}{c|}{NLO} & \multicolumn{2}{c|}{NNLO}
& \multicolumn{2}{c|}{NLO FFNS} & \multicolumn{2}{c|}{NNLO FFNS} \\
\hline
     0--0.0005   & $0$&$13 $ & $ 0$&$38 $   & $0$&$35 $ & $ 0$&$47 $   \\ 
0.0005--0.005    & $0$&$05 $ & $ 0$&$35 $   & $0$&$25 $ & $ 0$&$41 $  \\
 0.005--0.01     & $-0$&$11 $ & $ 0$&$13 $  & $-0$&$01 $ & $ 0$&$14 $  \\
  0.01--0.06     & $-0$&$15 $ & $-0$&$04 $  & $-0$&$10 $ & $-0$&$10 $  \\
  0.06--0.1      & $ 0$&$08 $ & $ 0$&$01 $  & $ 0$&$07 $ & $ 0$&$05 $  \\
   0.1--0.2      & $-0$&$12 $ & $-0$&$07 $  & $-0$&$15 $ & $-0$&$12 $  \\
   0.2--0.3      & $-0$&$16 $ & $-0$&$11 $  & $-0$&$21 $ & $-0$&$16 $  \\
   0.3--0.4      & $-0$&$20 $ & $-0$&$16 $  & $-0$&$23 $ & $-0$&$17 $  \\
   0.4--0.5      & $-0$&$09 $ & $-0$&$09 $  & $-0$&$10 $ & $-0$&$05 $  \\
   0.5--0.6      & $ 0$&$39 $ & $ 0$&$28 $  & $ 0$&$39 $ & $ 0$&$39 $  \\
   0.6--0.7      & $ 1$&$8  $ & $ 1$&$4  $  & $ 1$&$9  $ & $ 1$&$7  $  \\
   0.7--0.8      & $ 6$&$5  $ & $ 5$&$0  $  & $ 7$&$0  $ & $ 6$&$2  $  \\
   0.8--0.9      & $ 15$&$0 $ & $ 9$&$9 $   & $ 18$&$0 $ & $ 15$&$2 $  \\ \hline
\end{tabular}
\caption{\label{tab:t3}The values of the higher-twist coefficients $D_i$ of (\ref{HT}), in the chosen bins of $x$,
extracted from the NLO and NNLO GM-VFNS global fits and the NLO and NNLO FFNS 
fits to DIS data.}
\end{center}
\end{table}

Another difference in theoretical assumptions made when performing fit to 
data in order to extract PDFs is how to deal with the low $Q^2$ and low $W^2$
DIS data which is potentially susceptible to higher twist corrections to the
factorisation theorem. The majority of analyses choose a set of cuts which 
they deem to be large enough to eliminate the effect of higher twist effects,
and in the case of MSTW this is chosen to be $Q^2_{\min}=2\GeV^2$ and 
$W^2_{\min}=15\GeV^2$ (with the higher choice $W^2_{\min}=25\GeV^2$ 
for the small amount of 
$F_3(x,Q^2)$ data which is more likely to have large higher twist corrections)
where it has been checked in previous studies, e.g. 
\cite{Martin:2003sk}, that the PDFs and fit quality obtained are insensitive 
to smooth increases of the cuts in the upwards direction. However, some 
studies, e.g. \cite{Alekhin:2009ni} use lower cuts and parametrise the 
higher twist corrections as functions of $x$ and $Q^2$.

\begin{figure}[htb!]
\vspace{-1.3cm}
\centerline{\includegraphics[width=0.49\textwidth]{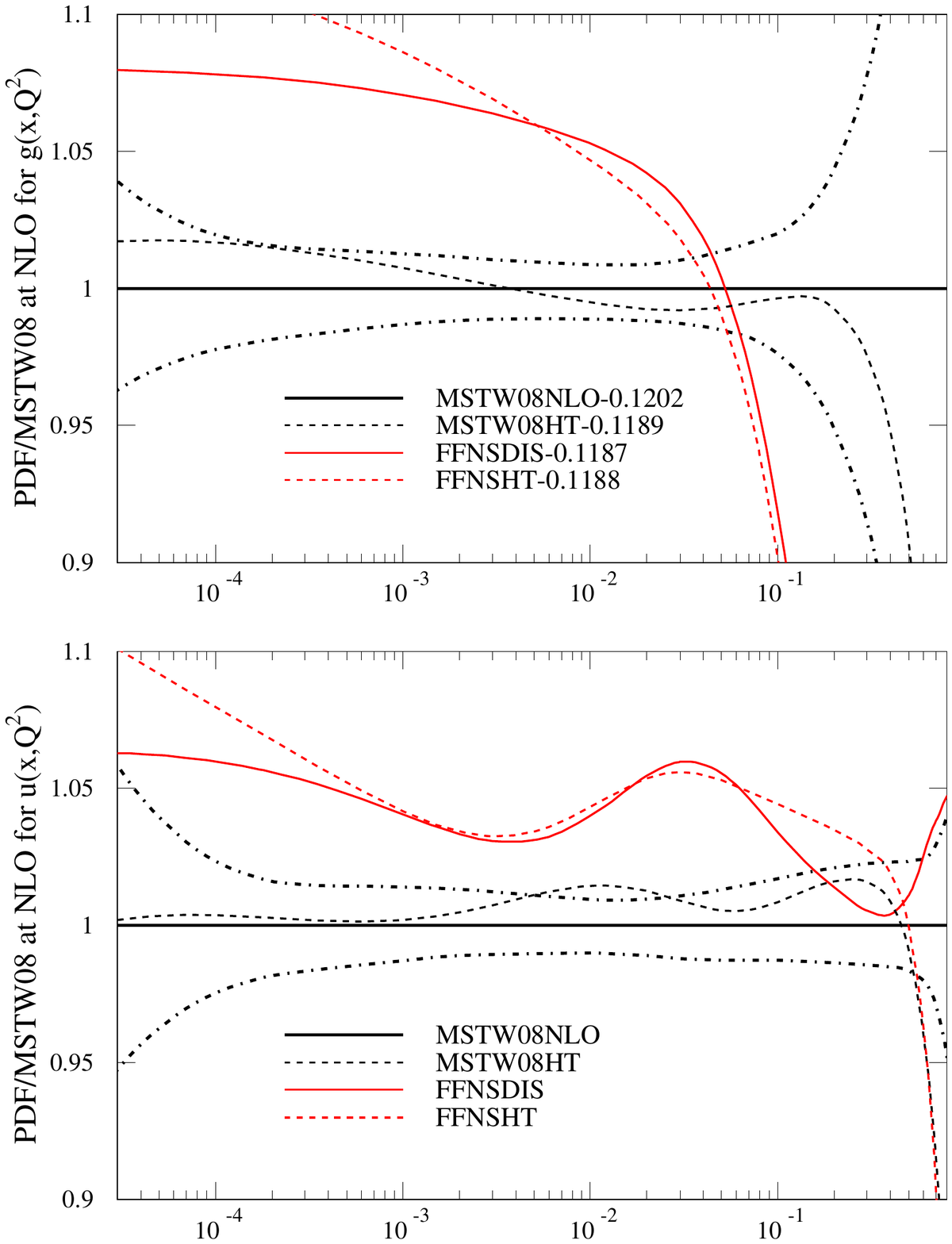}
\hspace{-1.5cm}\includegraphics[width=0.49\textwidth]{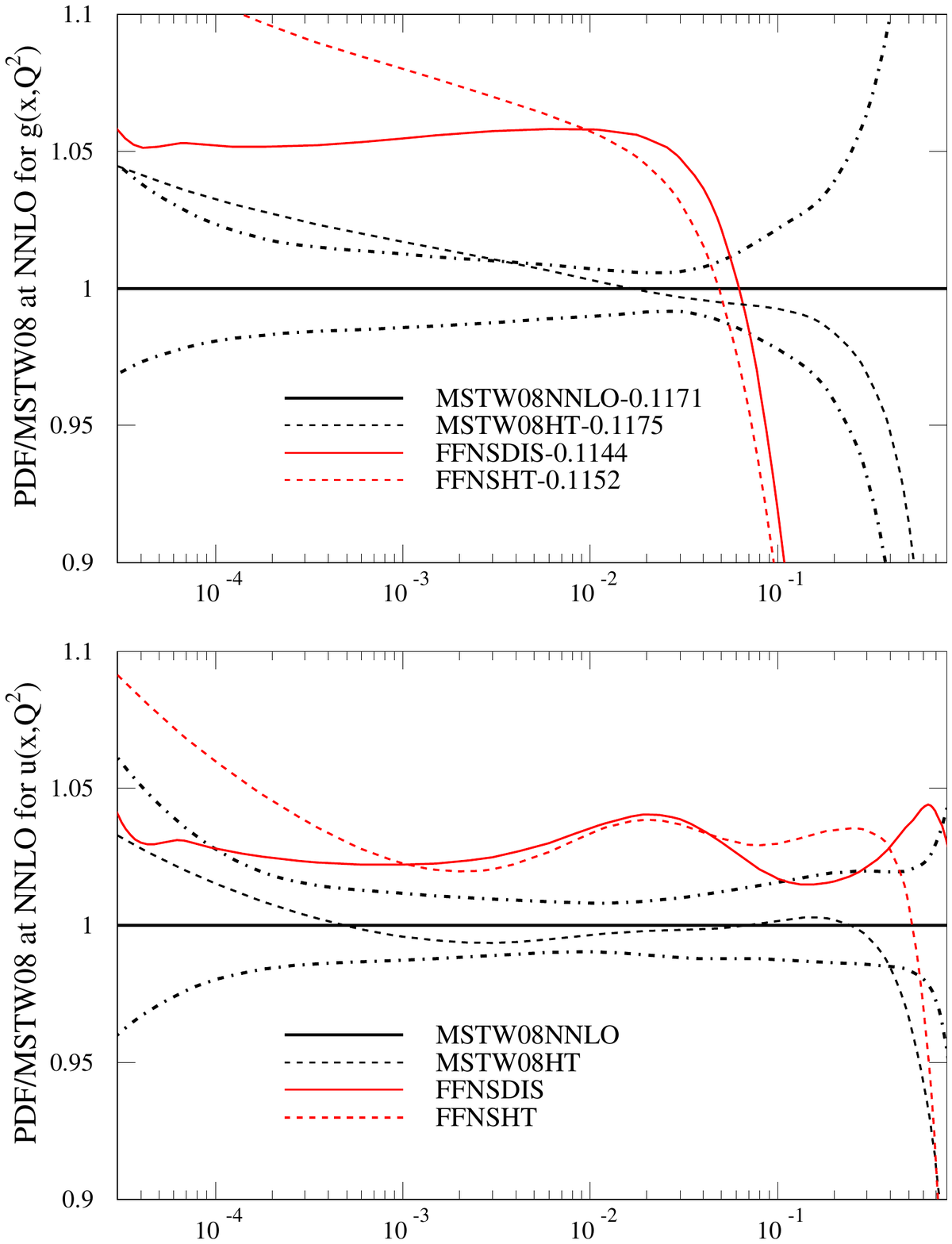}}
\vspace{-1cm}
\caption{Ratios of PDFs with higher twist corrections to PDFs without
at $Q^2=10,000\GeV^2$.}
\label{Fig8} 
\end{figure}

In order to check the sensitivity of the PDFs to this choice I have 
investigated the effect of lowering the $W^2$ cut for $F_2(x,Q^2)$ and 
$F_L(x,Q^2)$ to $5~\GeV^2$ (keeping that for $F_3(x,Q^2)$ unchanged) 
and parameterising higher twist corrections 
in the form $(D_i/Q^2)F_i(x,Q^2)$, where
\begin{equation} 
F_i(x,Q^2) =F^{\rm LT}_i(x,Q^2)\left( 1+\frac{D_i(x)}{Q^2}\right), \label{HT} 
\end{equation}
in 13 bins of $x$, and then fitting the 
$D_i$ and PDFs simultaneously, as in \cite{Martin:2003sk}. This is similar 
to the procedure in \cite{Alekhin:2009ni} and more recent PDF fits by 
the same group. It is less sophisticated than these fits, but the aim 
is simply to investigate the major changes in PDFs from including higher twist corrections, not to produce an official new set of PDFs. It is checked that 
results are insensitive to the treatment of longitudinal structure functions, which carry extremely little weight in the fit. The higher twist analysis
differs significantly from that in \cite{Ball:2013gsa} which took 
fixed higher twist parameterisations and kept the cuts of 
$Q^2_{\min}=3\GeV^2$ and $W^2_{\min}=12.5\GeV^2$ used as default 
by the NNPDF group, though variations, e.g. reversing the sign of the 
correction or doubling it were performed and the impact of these 
large changes investigated.  
The $D_i$ extracted in this study are shown in Table.~\ref{tab:t3}. They are 
similar to the older MRST study
in \cite{Martin:2003sk}, though larger at the smallest $x$. 
The effect on the PDFs and $\alpha_S(M_Z^2)$ compared to the 
default MSTW fit using GM-VFNS and all the same data sets is 
small, except for very high-$x$ quarks, 
as shown in Fig.~\ref{Fig8}. The value of $\alpha_S(M_Z^2)$ decreases slightly 
from 0.1202 to 0.1189 at NLO but actually increases slightly from 0.1171 
to 0.1175 at NNLO. The fit quality is shown at NLO in Table~\ref{tab:t4}
and at NNLO in Table~\ref{tab:t5}. The $\chi^2$ for the nuclear target 
structure function data is omitted here, as I will later consider a variety 
of fits where these data are left out. 

\begin{table}[h]
\begin{center}
\begin{tabular}{|l|c|c|c|c|} \hline
NLO &  &  & &\\
\hline 
& $\chi^2$ DIS & $\chi^2$ ftDY & $\chi^2$ jets & $\alpha_S^{n_f=5}(M_Z^2)$\\
& 2198pts  & 199pts & 186pts & \\
\hline
MSTW2008 HT   & 2077 & 233 & 164 & 0.1189 \\ 
MSTW2008 HT* (DIS+ftDY)  & 2045 & 222 & (201) & 0.1189 \\ 
 \hline
MSTW$n_f=3$ HT (DIS only) & 2060 &  &  ($>$300)& 0.1188\\ 
MSTW$n_f=3$ HT* (DIS only) & 2073 &  &  ($>$300) & 0.1175\\ 
MSTW$n_f=3$ HT* (DIS + ftDY) & 2075 & 237 & ($>$300)& 0.1179 \\ 
MSTW$n_f=3$ HT* (jets) & 2120 & 249 & 187 & 0.1199\\ 
MSTW$n_f=3$ HT* (jets+$Z$) & 2125 & 253 & 178 & 0.1215\\ 
\hline
MSTW$n_f=3$ HT* (DIS+ftDY)  & 2082 & 237 & 177 & 0.1200\\ 
\hline
\end{tabular}
\caption{\label{tab:t4} The $\chi^2$ values for DIS data, fixed target 
Drell Yan (ftDY) data and Tevatron jet data for various NLO fits performed using the 
GM-VFNS used in the MSTW 2008 global fit and using the $n_f=3$ FFNS 
for structure functions with reduced cuts and higher twist terms added.}
\end{center}
\end{table}

\begin{table}[h]
\begin{center}
\begin{tabular}{|l|c|c|c|c|} \hline
NNLO  & &  & &\\
\hline 
& $\chi^2$ DIS & $\chi^2$ ftDY & $\chi^2$ jets & $\alpha_S^{n_f=5}(M_Z^2)$\\
& 2198pts & 199pts & 186pts& \\
\hline
MSTW2008 HT   & 2039 & 241 & 175 & 0.1175 \\ 
MSTW2008 HT* (DIS+ftDY) & 2014 & 233 & (193) & 0.1175 \\ 
 \hline
MSTW$n_f=3$ HT (DIS only) & 2088 &  & ($>$300) & 0.1152\\ 
MSTW$n_f=3$ HT* (DIS only) & 2130 &  & ($>$300) & 0.1132\\ 
MSTW$n_f=3$ HT* (DIS + ftDY) & 2145 & 229 & ($>$300) & 0.1136 \\ 
MSTW$n_f=3$ HT* (jets) & 2174 & 246 & 183 & 0.1152\\ 
MSTW$n_f=3$ HT* (jets+$Z$) & 2179 & 253 & 173 & 0.1174\\ 
\hline
MSTW$n_f=3$ HT* (DIS+fyDY) & 2150 & 232 & ($>$300) & 0.1171\\ 
\hline
\end{tabular}
\caption{\label{tab:t5} The $\chi^2$ values for DIS data, fixed target 
Drell Yan (ftDY) data and Tevatron jet data for various NNLO fits performed using the 
GM-VFNS used in the MSTW 2008 global fit and using the $n_f=3$ FFNS 
for structure functions with reduced cuts and higher twist terms added.}
\end{center}
\end{table}

I have also repeated the higher twist study for fits using the 
FFNS for heavy flavour production, fitting to DIS data only. Again the 
results are shown in Fig.~\ref{Fig8}. The value of $\alpha_S(M_Z^2)$ 
only changes from  
from 0.1187 to 0.1188 at NLO and increases from 0.1144 
to 0.1152 at NNLO. The change in PDFs is fairly small and 
similar to that using the GM-VFNS and all global fit data. 
The extracted higher twist terms are shown in Table~\ref{tab:t3}. These
are similar to the GM-VFNS fit, but a little bigger, particularly NLO at 
small $x$. The  
fit quality is also shown at NLO in Table~\ref{tab:t4}
and at NNLO in Table~\ref{tab:t5}. There is less change in going from
GM-VFNS to FFNS when higher twist terms are included. In fact at NLO the FFNS 
DIS data only fit gives a slightly better fit to the DIS data than the full
higher twist MSTW2008 fit. However, this is no longer quite true for a DIS 
only GM-VFNS higher twist fit. However, the compatibility of the resultant 
PDFs with Tevatron jet data is far worse for the FFNS fit that the GM-VFNS 
fit.         

\begin{figure}[htb!]
\centerline{\includegraphics[width=0.49\textwidth]{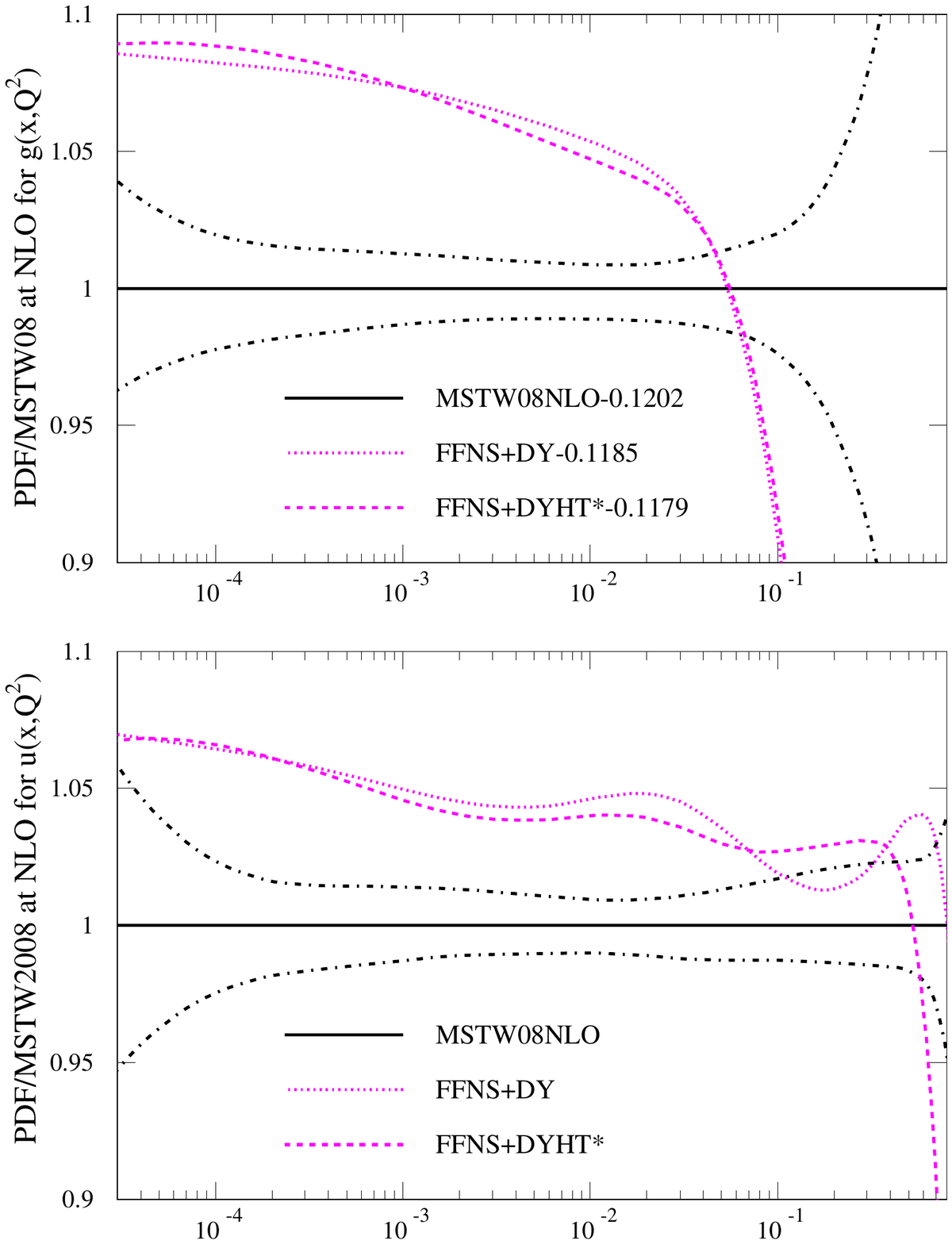}
\hspace{-1.5cm}\includegraphics[width=0.49\textwidth]{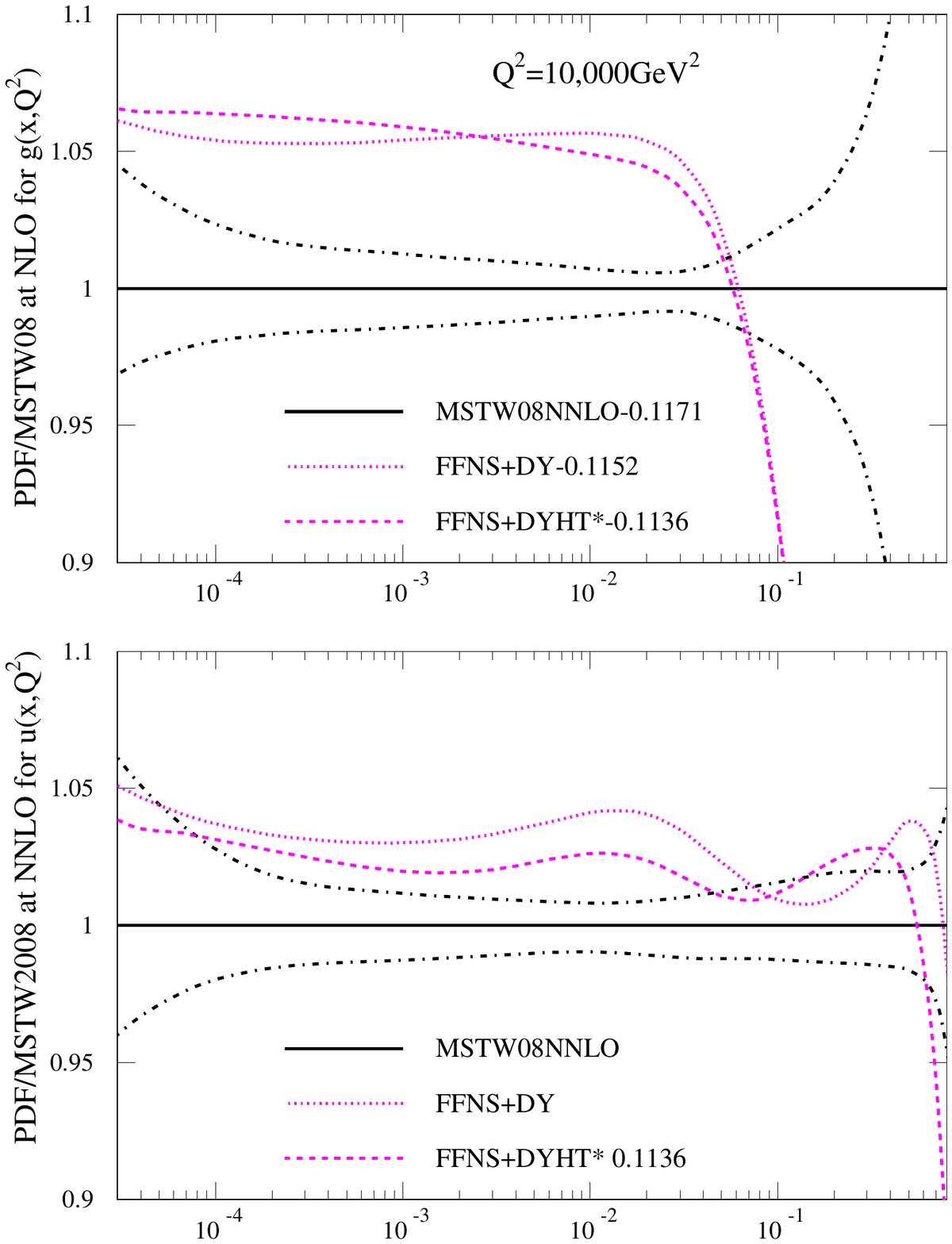}}
\vspace{-1cm}
\caption{Ratios of PDFs in two different FFNS fits to DIS plus Drell Yan 
data  to the MSTW2008 PDFs at $Q^2=10,000\GeV^2$.}
\label{Fig9} 
\end{figure}

\begin{figure}[htb!]
\centerline{\includegraphics[width=0.49\textwidth]{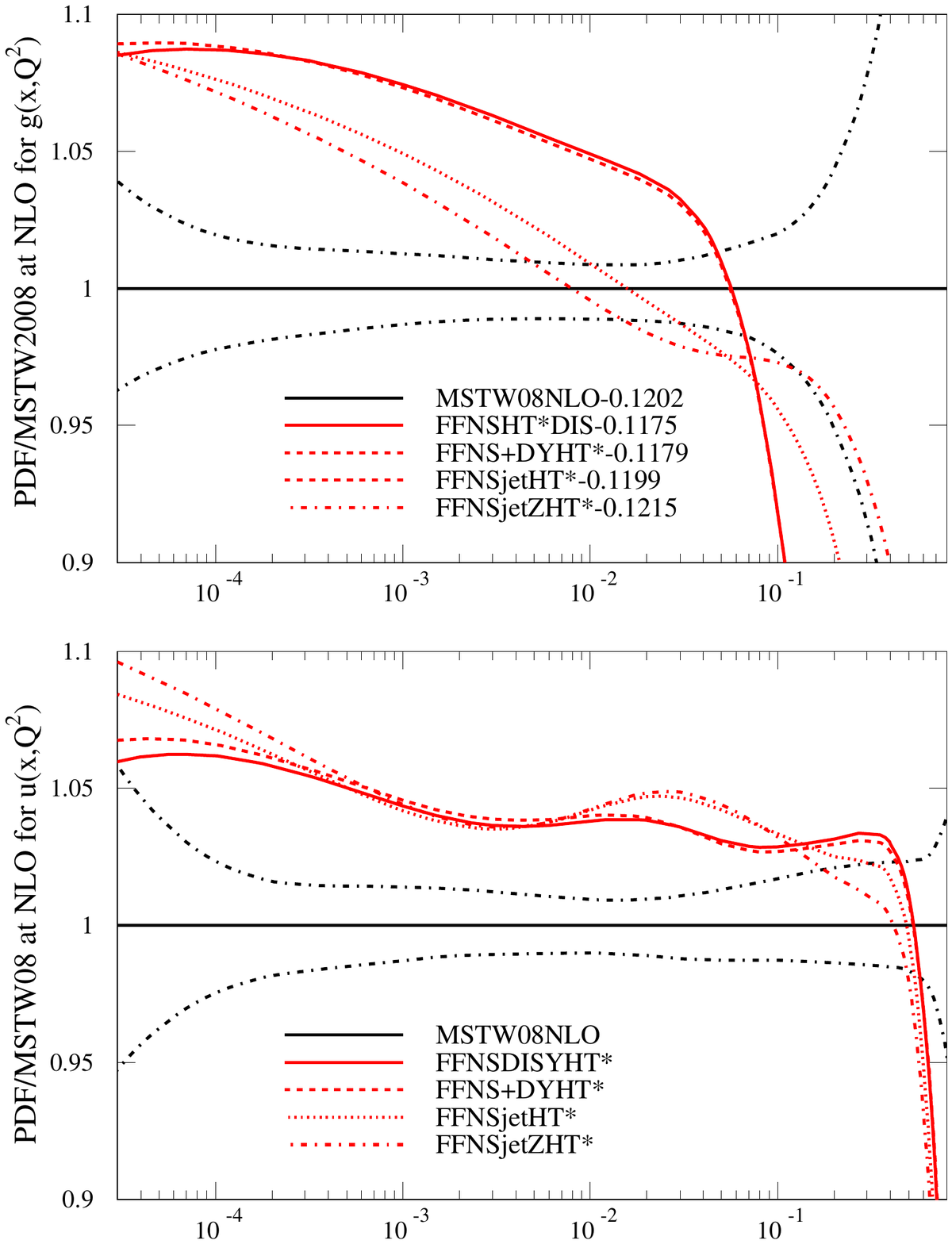}
\hspace{-1.5cm}\includegraphics[width=0.49\textwidth]{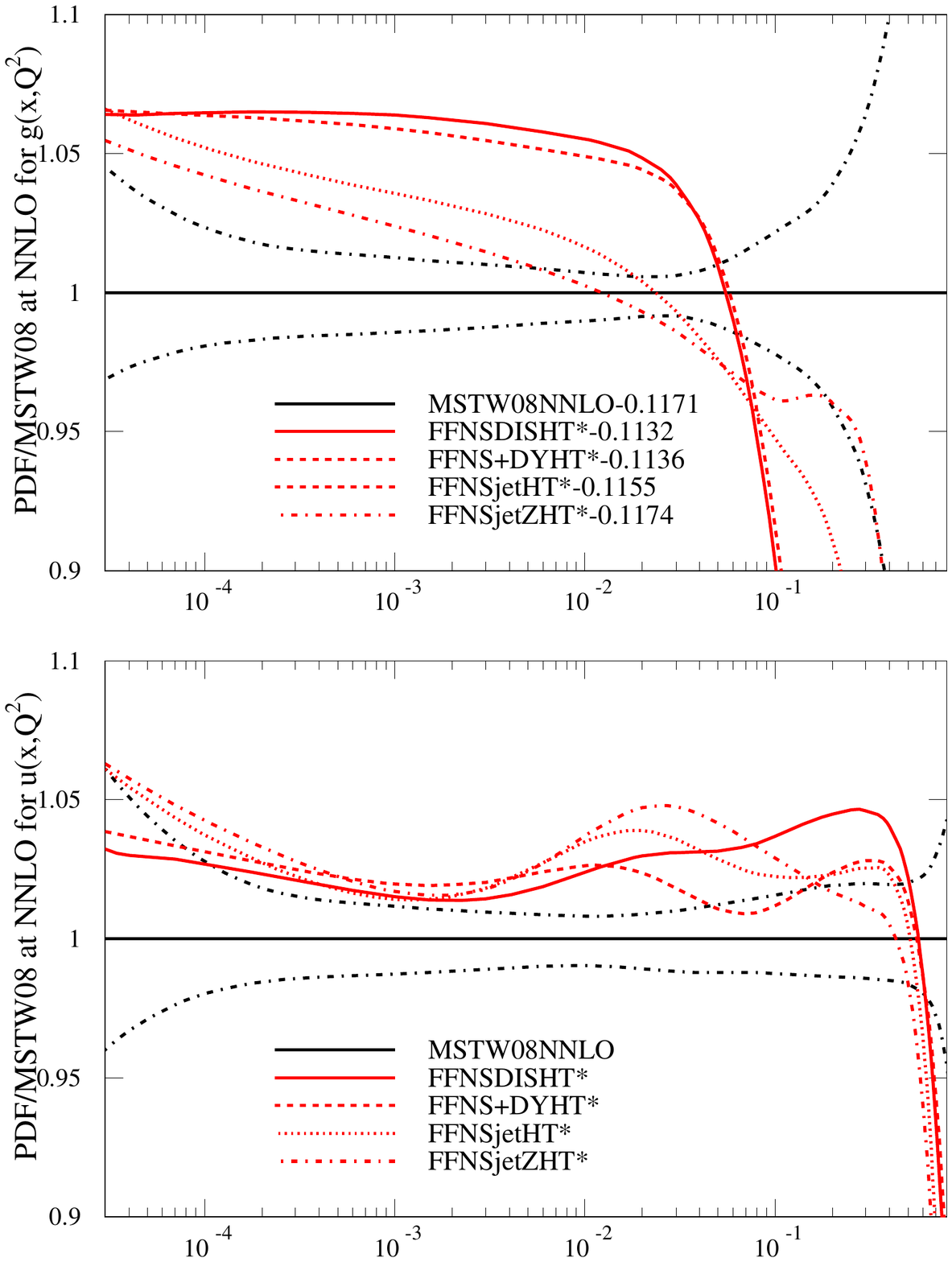}}
\vspace{-1cm}
\caption{Ratios of PDFs in various FFNS plus higher twist corrected fits 
to the MSTW2008 PDFs at $Q^2=10,000\GeV^2$. In the FFNS plus higher twist fits the nuclear target
inclusive DIS data is omitted and no higher twist corrections applied below 
$x=0.01$.}
\label{Fig10} 
\end{figure}

Although the value of $\alpha_S(M_Z^2)$ obtained from the FFNS fits with 
higher twist corrections is generally lower than that obtained in the GM-VFNS 
fits, particularly at NNLO, it is not as low as that obtained by other 
PDF groups which perform fits using the FFNS, e.g. 
\cite{Alekhin:2013nda,JimenezDelgado:2008hf}. In the latter of these 
there is sensitivity to the
input scale of the PDFs, with values of $Q_0^2$ lower than $1~\GeV^2$ 
leading to lower values of $\alpha_S(M_Z^2)$. I do not investigate this 
possibility since the MSTW PDF parameterisation is already such as to make the 
input gluon distribution rather different at any low scale. However, another 
difference in these fits compared to MSTW2008 is the absence of nuclear target
inclusive structure function data \cite{Tzanov:2005kr,Onengut:2005kv} which 
are dependent on nuclear corrections, but where the non-singlet 
$F_3(x,Q^2)$ data do favour high $\alpha_S(M_Z^2)$ values, as shown in 
\cite{Martin:2009iq}. Also in many higher twist studies the higher twist 
corrections are only included for $x>0.01$ 
Hence, I perform FFNS fits which restrict the  higher twist from the three 
lowest $x$ bins and simultaneously omit the less theoretically clean nuclear 
target data (except for dimuon cross sections, which constrain the strange quark). 
This results a series of fits labelled HT*. The fit quality for fits to 
only DIS data, DIS plus Drell Yan data and with the addition of Tevatron 
jet data and Tevatron $Z$ rapidity data is shown in 
Tables \ref{tab:t4} and \ref{tab:t5}. As mentioned earlier, in these tables 
the $\chi^2$ for DIS data does not include that for the nuclear target data, 
although the data has been included in the fits except for those labelled HT*. 
Removal of these data generally allow a slight improvement to the rest of 
the data, but this is compensated for by a (usually slightly larger) 
deterioration when the higher twist below $x=0.01$ is removed. As well as 
the FFNS fits I also show the fit quality for a GM-VFNS fit with 
$\alpha_S(M_Z^2)$ fixed to the same value as the full 
MSTW2008 higher twist fit, 
but the same data as the FFNS DIS plus Drell Yan fit is used. This is labelled 
MSTW2008HT*. For this approach the fit quality for the DIS plus Drell Yan data 
is the best exhibited, and the prediction for the Tevatron jets is quite 
good. The PDFs for the fits containing DIS 
plus fixed target Drell Yan data are compared to MSTW2008 for two variants 
of the FFNS fit in Fig.~\ref{Fig9} and 
the full range of HT* fits are shown in Fig.~\ref{Fig10}. The additional 
changes in the HT* fits do result in slightly lower values of $\alpha_S(M_Z^2)$, particularly 
at NNLO, with values of $\alpha_S$ of $\alpha_S(M_Z^2)=0.1179$ at NLO and 
$\alpha_S(M_Z^2)=0.1136$ at NNLO for the fits without Tevatron data. These 
are very close to those in \cite{Alekhin:2009ni}, where the FFNS scheme 
choice, data types, and form of higher twist (and the resulting PDFs) 
are similar. The change in the PDFs in going from the FFNS fits to FFNSHT* 
fits is not large at all, as seen in Fig.~\ref{Fig9}, 
with the essential features of the differences
between FFNS and GM-VFNS PDFs being fully maintained. 

I have also made some further checks on the general validity of the results. 
It was noted in \cite{Martin:2010db} that when using the default GM-VFNS for
the MSTW2008 fit the best fit quality was obtained for values of the pole mass
$m_c$ different to the default $m_c=1.4\GeV$. At NLO the global $\chi^2$ could 
decrease by just a couple of units with a very slightly larger value
$m_c=1.45~\GeV$, but at NNLO the global $\chi^2$ could 
decrease by 24 units if the lower value of $m_c=1.26~\GeV$ is used. In the 
FFNS fits a very slight decrease in $\chi^2$ of a few units is obtained 
at NLO 
if $m_c$ lowers by $0.1~\GeV$ or less and at NNLO an improvement in $\chi^2$ 
of up to 30 units can be achieved for $m_c=1.2$-$1.25~\GeV$. Hence, the 
improvements in fit quality possible using the GM-VFNS and FFNS are very 
similar, perhaps marginally better for FFNS, and FFNS prefers a slight lower 
optimum $m_c$ value. None of this has any significant effect on the relative
differences in PDFs or $\alpha_S(M_Z^2)$. Also, as demonstrated in Section 3,
the differences between FFNS and GM-VFNS can be very 
largely understood in terms 
of the leading $\ln(Q^2/m_c^2)$ terms in the perturbative expansions. 
These are completely unaltered by a change in quark mass scheme of 
$m_c \to m_c(1+c \alpha_S + \cdots)$. Indeed, there is only a fairly minor 
change in PDFs from \cite{Alekhin:2009ni} to \cite{Alekhin:2012ig}, and 
almost no change in $\alpha_S(M_Z^2)$, despite the change from the pole mass 
to $\msbar$ mass schemes. Perhaps the most striking change, an increase 
in sea quarks near $x=0.01$ is due to the inclusion of the combined HERA data
\cite{Aaron:2009aa}, an effect noticed elsewhere, e.g. \cite{Thorne:2010kj}. 
As a final check, fits were performed using approximations to 
the full NNLO heavy flavour DIS coefficients. Wider variations in 
coefficient functions were allowed than options $A$ and $B$ in 
\cite{Kawamura:2012cr}. At best the NNLO 
FFNS fits improved quality by about 40-50 units - significant but still 
leaving them some way from the GM-VFNS fit quality at NNLO. The change in PDFs 
and $\alpha_S(M_Z^2)$ is never 
very large, and the very best fits actually preferred a marginally lower 
$\alpha_S(M_Z^2)$ value. Hence, the conclusions on fit quality, 
the PDF shape and $\alpha_S(M_Z^2)$ values are stable under a variety of 
variation in the full details of the fit. The general features of the 
FFNS fits producing gluon distributions which are about $10\%$ lower at 
$x\sim 0.1$ at $Q^2=10,000~\GeV^2$ than when using GM-VFNS, but rising to $5\%$
(or more) greater below $x=0.01$, along with a light quark distribution which 
is a few percent bigger at most $x$ values seems to be largely insensitive to 
any other variations in procedure or data fit. The reduction of 
$\alpha_S(M_Z^2)$ also seems to be a stable feature, but the precise 
difference is more sensitive to details of the fit.

\section{Fixed Coupling}

\begin{figure}[htb!]
\centerline{\includegraphics[width=0.49\textwidth]{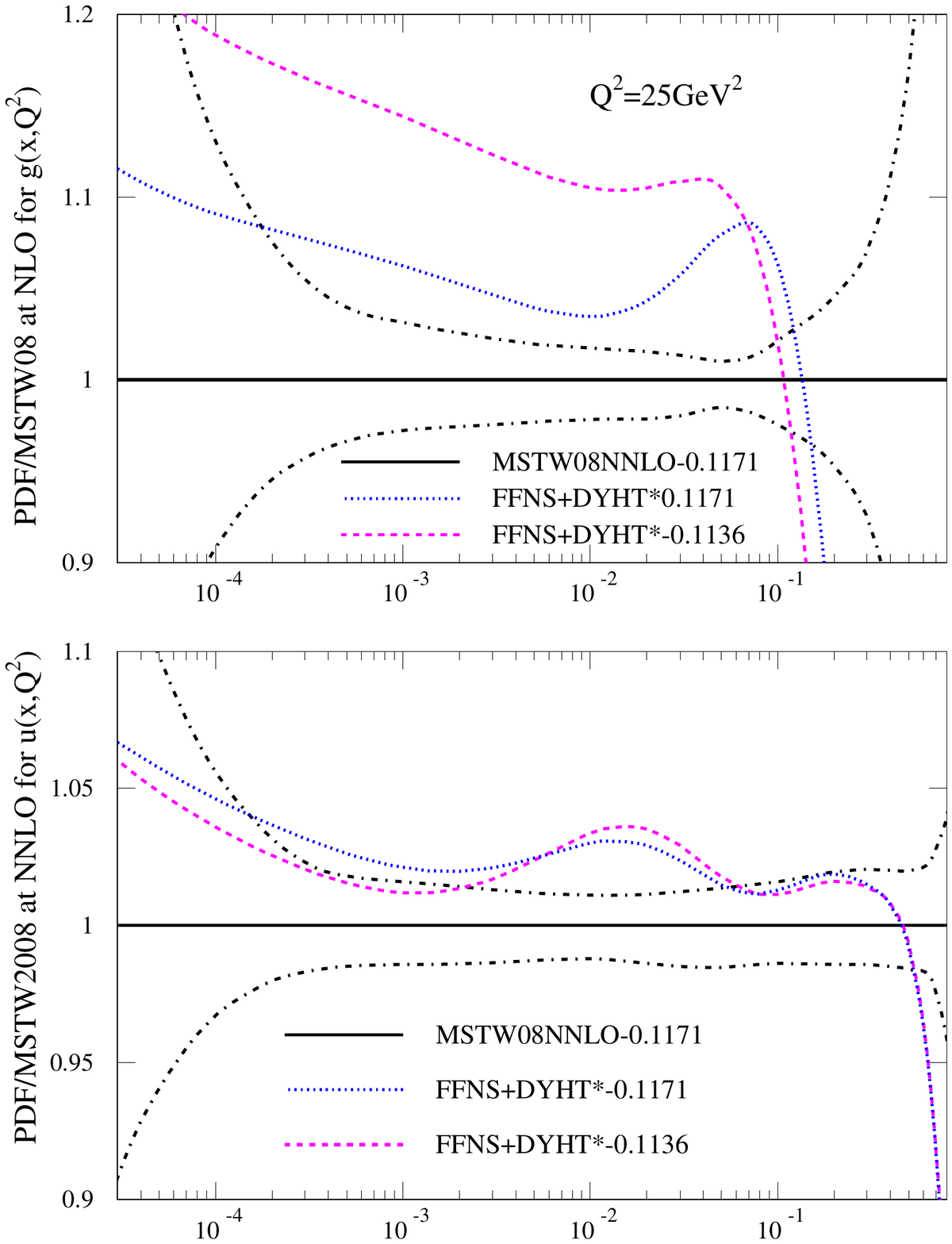}
\hspace{-1.5cm}\includegraphics[width=0.49\textwidth]{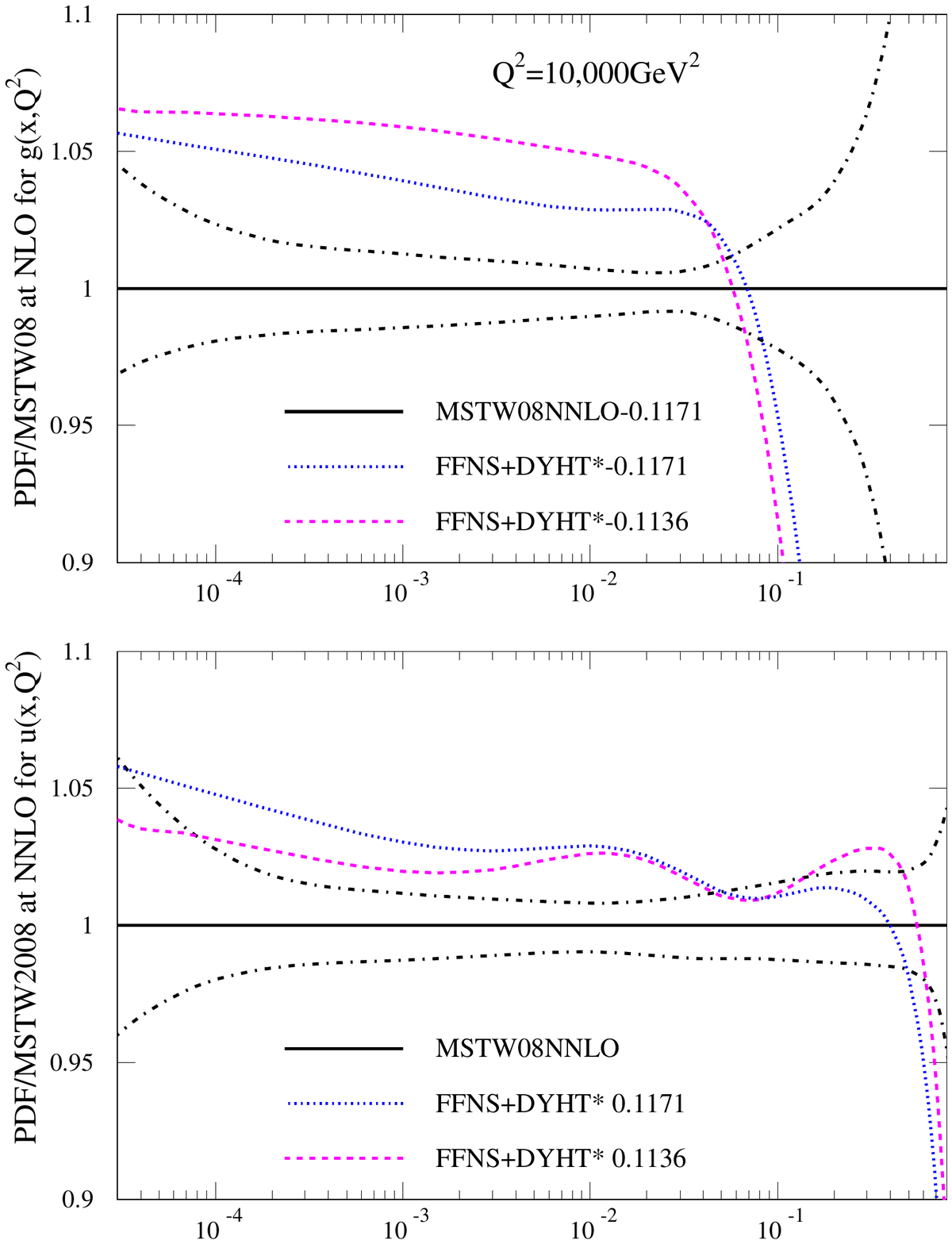}}
\vspace{-1cm}
\caption{The ratio of FFNS PDFs from NNLO 
fits with both free (red) and fixed $\alpha_S(M_Z^2)$ (blue)
to the MSTW2008 PDFs at $25~\GeV^2$ (left) and at $10,000~\GeV^2$ (right). }
\vspace{-0.4cm}
\label{Fig11} 
\end{figure}

Finally, in order to investigate why the value of $\alpha_S(M_Z^2)$ obtained
in FFNS fits is lower than in GM-VFNS fits I  
also perform a NNLO fit to DIS and low-energy DY data 
where $\alpha_S(M_Z^2)$ 
is fixed to the higher value obtained in the GM-VFNS. 
I also perform a fit with $\alpha_S(M_Z^2)=0.120$ at NLO, though the 
relative change in the coupling is less significant at NLO. This fixed 
coupling results in the  
FFNS gluon being a little closer to that using GM-VFNS, as shown at NNLO
in Fig.~\ref{Fig11} for $Q^2=25~\GeV^2$ and $Q^2=10000~\GeV^2$,  
and very similar to the gluon in \cite{Ball:2013gsa}, where studies are 
performed with 
fixed $\alpha_S(M_Z^2)$. There is little change in the light quarks
in the FFNS fit when the coupling is held fixed. 
The fit quality is shown in Tables \ref{tab:t4} and \ref{tab:t5} 
The FFNS fit is 8 units worse when $\alpha_S(M_Z^2)=0.1171$
than for 0.1136. (The deterioration at NLO is very slightly less.) 
The fit to HERA data is better, but it is worse for fixed  
target data.

By examining the change in the gluon in the FFNS fit when $\alpha_S(M_Z^2)$ 
is fixed one can understand the need for $\alpha_S$ to be smaller in 
FFNS. To compensate for smaller $F_2^c(x,Q^2)$ at 
$x \sim 0.05$ the FFNS gluon must be bigger in this region, and 
from the momentum sum rule, is therefore smaller at 
high $x$. The correlation between the high-$x$ gluon and 
$\alpha_S(M_Z^2)$ when fitting high-$x$ fixed target 
DIS data drives $\alpha_S$ down
(for reduced gluon the quarks fall with $Q^2$ more quickly, 
hence the need to  lower $\alpha_S$ to slow evolution),
requiring  the small $x$ gluon to even bigger. As the fit undergoes 
iterations this pattern is repeated until the best fit is reached with a 
lower $\alpha_S(M_Z^2)$ value and significantly modified gluon shape.

\section{Conclusions}

In this article I have investigated whether the different theoretical 
choices in fits 
to data in order to determine partons distribution functions (PDFs) can 
influence the PDFs, the value of $\alpha_S(M_Z^2)$ and the fit quality. I 
come to the strong conclusion that within the context of the MSTW2008 
global fit the choice of a FFNS for heavy flavour 
production in deep inelastic scattering, as opposed to a GM-VFNS, leads to a 
lower $\alpha_S(M_Z^2)$, a gluon distribution which is much lower at very 
high-$x$ but smaller at small $x$, and larger light quarks over most $x$ 
values. In contrast, making the $Q^2$ and $W^2$ cuts on the data less 
conservative and introducing higher twist corrections which are fit to the 
data makes little difference to PDFs, except at very high $x$ and also 
little difference to $\alpha_S(M_Z^2)$, particularly at NNLO. 

This result concerning the importance of the choice of heavy flavour scheme 
used might seem surprising. It is known that the FFNS and a well-defined 
GM-VFNS will converge towards each other as the perturbative order is 
increased. At higher orders more and more large logs in $Q^2/m_c^2$ are 
included in the FFNS and 
the ambiguities in the GM-VFNS definition near threshold are shifted to 
higher and higher order. Indeed, it has often been suggested, 
e.g. \cite{Botje:2011sn}, that the omission of Tevatron jet data is the 
likely source of the smallness of the high-$x$ gluon in some PDF sets.
This is undoubtedly partially true. It is seen in Fig.~\ref{Fig3} of this 
article that when fitting using FFNS the inclusion of jet data raises the 
gluon for $x>0.1$ and $\alpha_S(M_Z^2)$ (in \cite{Alekhin:2012ig} top
pair production cross sections are raised when Tevatron jet data is 
included). However, GM-VFNS fits without jet data do not automatically have a 
lower high-$x$ gluon or $\alpha_s(M_Z^2)$ value - it is simply that 
constraints on both are loosened. For example, it is not really 
clear why for the 
HERAPDF1.5 PDFs in \cite{CooperSarkar:2011aa}, which fit HERA DIS data only, 
the NNLO high-$x$ gluon is harder than NLO. Hence, the inclusion of jet data
or not is only part of reason for significant PDF differences. It has also 
been argued, e.g. \cite{Alekhin:2013nda}, that it is the absence of NNLO 
corrections to jet production that leads to differences in the gluon in 
different PDF sets at NNLO, i.e. the NNLO high-$x$ gluon is 
being overestimated due 
to missing positive NNLO corrections. I find this unconvincing. In the 
MSTW2008 fits threshold corrections of $\sim 20\%$ from \cite{Kidonakis:2000gi}
are used in NNLO fits. It was shown recently \cite{Kumar:2013hia}
that the absence of jet radius
$R$ dependence in these terms leads to an underestimate of the full NLO result
in the threshold approximation of \cite{Kidonakis:2000gi}. However, improved 
threshold calculations in \cite{deFlorian:2013qia} shown little $R$ 
dependence at NNLO, and the size of corrections at NNLO inferred from 
\cite{deFlorian:2013qia} is quite 
similar to that used in MSTW fits. Additionally, in 
\cite{Watt:2013oha} extreme changes in the assumed NNLO corrections for 
Tevatron jets are considered and changes in PDFs and $\alpha_S(M_Z^2)$ are 
considerably smaller than those seen from changing the flavour scheme in 
this article. Hopefully a full NNLO calculation of jet cross sections    
\cite{gggp,Currie:2013dwa} will settle this dispute soon. 
Furthermore, the issue of NNLO jet cross sections only affects NNLO PDFs, and 
the general features of the differences between different PDF sets are
all very similar at NLO and at NNLO, so attributing them to effects unique 
to NNLO seems rather unlikely to be correct. 

In fact the study in this article
began at NLO 
in \cite{Thorne:2012az}, where significant differences between FFNS and 
GM-VFNS was seen. 
As well as building on the phenomenological results of 
this initial study by showing a similar effect is indeed present at NNLO, 
and is consistent with results comparing FFNS and GM-VFNS in 
\cite{CooperSarkar:2007ny} and \cite{Ball:2013gsa}, this article shows 
exactly why this effect exists by studying the form of the leading logarithmic
contribution to $(d\,F_2^c(x,Q^2)/d\,\ln Q^2)$ in FFNS and GM-VFNS. It
is shown in Section 3 that one can understand 
exactly why evolution at high $Q^2$ is considerably 
slower in FFNS than in GM-VFNS for $x\sim 0.05$, and that the difference 
between the two will only converge at very high perturbative order. This
has an important impact on the fit to inclusive DIS data since there is a very
large amount of $F_2(x,Q^2)$ 
HERA data at high $Q^2$ for $0.1<x<0.01$, and $F_2^c(x,Q^2)$
is a large contribution to this. Since the charm contribution in FFNS is 
lower at high-$Q^2$ it is clear that light quarks will be higher to 
compensate. The change in the gluon and $\alpha_S(M_Z^2)$ is less obvious, 
but an argument for their form is put forward in Section 5. 

Hence, I conclude that the use of GM-VFNS and FFNS will result in 
significantly different PDFs and $\alpha_S(M_Z^2)$ up to NNLO, whereas higher 
twist corrections are not important so long as their absence is accompanied 
by sufficiently high cuts on $W^2$ and $Q^2$. The difference 
between FFNS and GM-VFNS PDFs will be 
moderated as the fit becomes more global and more data types are added, 
but the fit quality seems to be better using a GM-VFNS and less tension 
between different data sets is observed. Indeed, PDFs which are obtained using 
a GM-VFNS are already seen to match LHC jet data very well 
\cite{Ball:2012cx,Watt:2013oha}. Additionally, one may feel that if there
is slow convergence of a expansion which contains finite orders of
$\alpha_S^n\ln^n(Q^2/m_c^2)$ to the result of a 
fully resummed series of these terms then it is
theoretically preferable to use the latter. Therefore, I advocate the use 
of a GM-VFNS in PDF fits to data.

\vspace{0.3cm}

\section*{Acknowledgements}

I would like to thank A. D. Martin, W. J. Stirling  and G. Watt for
numerous discussions on PDFs, and A.M Cooper Sarkar, S. Forte, P. Nadolsky
and J. Rojo for discussions on heavy quarks schemes. This work is
supported partly by the London Centre for Terauniverse Studies (LCTS),
using funding
from the European Research Council via the Advanced Investigator Grant 267352.
I would also like to thank the Science and Technology Facilities Council 
(STFC) for support.

\end{document}